\documentclass[superscriptaddress, amsmath, amssymb, aps, twocolumn, longbibliography]{revtex4-2}
\usepackage{graphicx}
\usepackage{dcolumn}
\usepackage{amsthm}
\usepackage{bm}
\newcommand{\ket}[1]{|#1\rangle}
\newcommand{\bra}[1]{\langle#1|}

\newcommand{\Tr}[1]{\textrm{Tr}(#1)}

\pdfoutput=1
\usepackage{graphicx,graphics,epsfig,subfigure,times,bm,bbm,amssymb,amsmath,amsthm,mathrsfs,MnSymbol}
\usepackage{gensymb}
\usepackage{amsfonts}
\usepackage[matrix,frame,arrow]{xypic}
\usepackage[pdfstartview=FitH]{hyperref}
\hypersetup{
	colorlinks=true,       
	linkcolor=red,          
	citecolor=magenta,        
	filecolor=magenta,      
	urlcolor=cyan,           
	runcolor=cyan}
\usepackage{epstopdf}
\usepackage[pdftex]{color}
\usepackage{braket}
\usepackage{enumerate}
\usepackage[normalem]{ulem}
\usepackage[usenames,dvipsnames]{xcolor}
\usepackage{multirow}
\usepackage{mathtools}

\definecolor{orange}{rgb}{1,0.5,0}

\newcommand{\ignore}[1]{}
\usepackage{geometry}

\newcommand{\n}{\nonumber\\}
\newcommand{\ex}[1]{\langle #1\rangle}
\newcommand{\Supp}[1]{{\rm Supp}(#1)}

\ignore{
	\documentclass[eprintnumbers,amsmath,amssymb,onecolumn,a4paper,caption ]{article}
	\usepackage{amsfonts}
	\usepackage{amssymb}
	\usepackage{mathrsfs}
	\usepackage{mathbbold}
	\usepackage{bbm}
	\usepackage{mathrsfs}
	\usepackage{dcolumn}
	\usepackage{bm}
	\usepackage{times,epsfig,amssymb,amsmath}
	\usepackage{float}
	\usepackage{subfigure}
	\usepackage{geometry}\geometry{left=2.5cm,right=2.5cm,top=3cm,bottom=3cm}

	\usepackage{color}

}

\usepackage{changes}
\definechangesauthor[name={Yun-Hao Shi},color=blue]{syh}

\begin{document}

\title{Quantum Estimation with State Symmetry-Induced Optimal  Measurements}

\author{Jia-Xuan Liu}
\thanks{These authors contributed equally to this work.}
\affiliation{Hefei National Research Center for Physical Sciences at the Microscale
and School of Physical Sciences, Department of Modern Physics, University
of Science and Technology of China, Hefei, Anhui 230026, China}

\author{Hai-Long Shi}
\thanks{These authors contributed equally to this work.}
\affiliation{QSTAR, INO-CNR, and LENS, Largo Enrico Fermi 2, 50125 Firenze, Italy}

\author{Chunfeng Wu}
\email{chunfeng_wu@sutd.edu.sg}
\affiliation{Science, Mathematics and Technology, Singapore University of Technology and Design, 8 Somapah Road, Singapore 487372, Singapore}

\author{Sixia Yu}
\email{yusixia@ustc.edu.cn}
\affiliation{Hefei National Research Center for Physical Sciences at the Microscale
and School of Physical Sciences, Department of Modern Physics, University
of Science and Technology of China, Hefei, Anhui 230026, China}
\affiliation{Hefei National Laboratory, University of Science and Technology of
China, Hefei 230088, China}

\begin{abstract}
A central challenge in quantum metrology is identifying optimal measurements that saturate the quantum Cram\' er–Rao bound under realistic constraints, e.g., local measurements.
We show that symmetries of the probe state provide a general principle for identifying optimal measurement strategies. Building on this idea, we demonstrate that when a parameter is encoded in the real coefficients of a fixed-basis expansion, the optimal measurement reduces to projection in that basis, with an application to critical metrology.
Under local-measurement constraints, we show that local state symmetries provide a systematic route to constructing optimal local measurements. We illustrate this framework using graph states, explicitly constructing optimal local measurements from their local symmetries. Furthermore, weak and strong connection rules are introduced to  generate broader classes of graph states that achieve Heisenberg-scaling precision using local measurements.
By relaxing the number of stabilizer generators, graph states are extended to a stabilizer-code subspace.
Analytical and numerical results show that coherent states in these subspaces offer multiple metrological advantages: high precision, partial noise resilience, local-measurement accessibility, and built-in error correction.
These findings advance the theory of optimal measurements in quantum metrology and underscore the central role of state symmetry.
\end{abstract}
\pacs{Valid PACS appear here}
\maketitle

\noindent
\textbf{\large{Introduction}}

\noindent 
High-precision parameter estimation lies at the heart of quantum metrology, with both fundamental and practical significance \cite{PhysRevD.23.1693,doi:10.1126/science.1104149,Giovannetti2011,Toth_2014,RevModPhys.90.035005,RevModPhys.89.035002,PhysRevA.77.012317,PhysRevB.109.L041301,PhysRevLett.100.220501,PhysRevX.8.021022,PhysRevLett.107.083601,shi2025quantum,PhysRevLett.132.100803,PhysRevLett.124.110502,PhysRevLett.132.210801,jkjj-3gvb,h6sr-yxgw,montenegro2025quantum,DiFresco2024metrology,Zhang2025,10.21468/SciPostPhys.13.4.077,DeMille2024,zhou2020saturating}. While global measurements can, in principle, saturate the ultimate quantum limits, their implementation becomes experimentally challenging as system size increases \cite{Friis2017}. Local measurements, by contrast, are more accessible in multipartite scenarios but are generally believed to fall short of saturating ultimate quantum limits, except in highly symmetric cases \cite{PhyRevX8021022,PhysRevA.111.022436,f2jf-bg7g}. 
A notable example is the Greenberger–Horne–Zeilinger (GHZ) state \cite{10.1119}, a paradigmatic multipartite entangled state. 
When used to estimate a phase parameter $\theta$, the variance of an unbiased estimator scales as $1/N^2$ under local measurements, where $N$ denotes the number of qubits.
This scaling saturates the Heisenberg limit (HL), offering an $O(N)$ improvement over protocols that do not exploit entanglement \cite{PhysRevLett.96.010401}.
These observations naturally raise a fundamental question: which structural features of multipartite states, such as GHZ states, enable HL precision under local measurements, and can these features be systematically extended to broader families of states?

Yet, in realistic settings, environmental noise inevitably destroys quantum coherence and entanglement \cite{RevModPhys.89.041003,PhysRevLett.123.180504,PhysRevA.100.012308,wfyl-wtz3}, reducing achievable sensing precision \cite{PhysRevLett.79.3865,Escher2011,Demkowicz-Dobrzanski2012,PhysRevLett.113.250801}. Particularly in the noisy intermediate-scale quantum (NISQ) era \cite{Preskill2018quantumcomputingin,RevModPhys.94.015004}, strategies such as quantum error correction \cite{PhysRevLett.112.080801,PhysRevLett.112.150802,Lu2015,Zhou2018,kurzyna2025microwavefieldquantummetrologyerror,kwon2025virtualpurificationcomplementsquantum}, dynamical decoupling \cite{Taylor2008,Sekatski_2016,PhysRevB.86.045214}, time optimization \cite{PhysRevLett.109.233601,PhysRevA.108.022413}, and feedback control \cite{Hirose2016,PhysRevA.91.033805,PhysRevA.96.012117,PhysRevLett.130.240803} have been proposed to mitigate the effects of noise. 
These considerations naturally extend the earlier question:  can HL precision still be achieved under local measurements in the presence of realistic noise?

We address these challenges by establishing a general theoretical framework that links state symmetries to optimal measurement strategies (Theorems 1 and 2).
Theorem 3 further demonstrates that, under local-measurement constraints, local state symmetries can be used to construct optimal local measurements.
We apply this framework to graph states, explicitly deriving optimal local-measurement protocols from their local symmetries.
To broaden the class of useful resources, we introduce weak and strong connection rules, which generate extended families of graph states capable of achieving HL scaling with local measurements. 
By reducing the number of stabilizer generators, we define a relaxed-stabilizer subspace and show that coherent states within this subspace exhibit enhanced noise resilience compared with the standard GHZ state.
Because the relaxed-stabilizer subspace can be viewed as a stabilizer-code subspace, quantum error correction can be naturally incorporated to further mitigate errors.
In this work, we identify state symmetries as a unifying principle connecting precision, noise robustness, and error-correcting capability, and demonstrate that exploiting these symmetries enables optimal measurements for quantum metrology using only local operations.

~\\

\begin{figure}[t]
\begin{centering}
\includegraphics[scale=0.19]{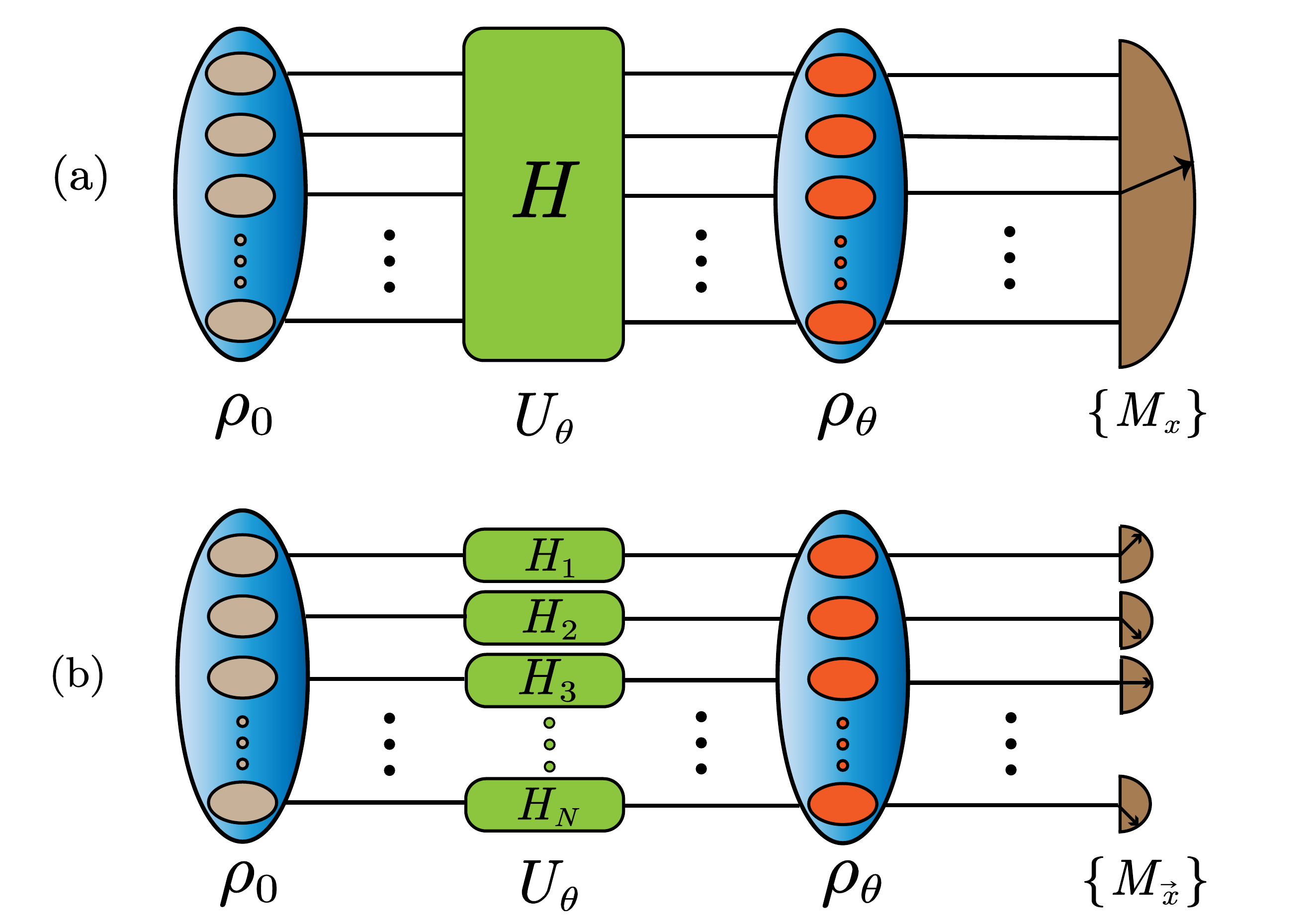}
\par\end{centering}
\caption{\label{fig:1}
Schematic illustration of protocols for estimating a global phase parameter $\theta$ in an $N$-qubit system. 
(a) Global unitary encoding $U_\theta=\exp(-i\theta H/2)$ generated by an arbitrary Hamiltonian $H$, followed by collective measurements $\{M_x\}_{x\in \mathcal X}$. 
(b) Local unitary encoding $U_\theta=\exp(-i\theta H/2)$ generated by a linear Hamiltonian $H=\sum_j H_j$, followed by local measurements $\{M_{\vec x}=\bigotimes_{j=1}^N M_{x_j}\}_{\vec x\in\mathcal X}$.} 
\end{figure}

\noindent
\textbf{\large{Results}}
~\\
\noindent  
\textbf{State Symmetry-induced Optimal Measurements}

\noindent 
A pure state $|\psi\rangle$ is said to exhibit a state symmetry with respect to a Hermitian operator $S$ if it satisfies  $S |\psi\rangle = |\psi\rangle$.
This invariance implies that the action of $S$ leaves $|\psi\rangle$ unchanged, a property that can be exploited in constructing optimal measurements.
For a mixed state $\rho_0 = \sum_j p_j |\psi_j\rangle\langle \psi_j|$, the definition extends naturally by requiring that every  eigenvector within its support remains invariant under $S$,
\begin{align}\label{State_Symmetry}
S \ket{\psi_j} =\ket{\psi_j}, \quad \forall \ket{\psi_j}\in \Supp{\rho_0},
\end{align}
where $\Supp{\rho_0}$ denotes the support of $\rho_0$, i.e., the subspace spanned by its eigenvectors with nonzero eigenvalues.

Having established the concept of state symmetry, we now consider its application in quantum metrology, where such a symmetric state $\rho_0$ serves as the initial probe, as depicted in Fig.~\ref{fig:1}(a).  
Specifically, the probe state  $\rho_0$ undergoes a phase-encoding process governed by a unitary evolution
\begin{align}\label{encoded_state}
\rho_\theta = U_\theta \rho U_\theta^\dagger, \quad U_\theta = \exp(-i \theta H / 2),
\end{align}
where $H$ is the Hamiltonian generating the evolution and $\theta$ is the parameter to be estimated.
To infer $\theta$, we perform a quantum measurement described by a positive operator-valued measure (POVM)  $\{M_x\}_{x\in\mathcal X}$, yielding outcomes $x$ with probability $q(x|\theta) = \Tr{M_x \rho_\theta}$. 

The Cram\'{e}r–Rao bound~\cite{Rao1945,Cramer1946} provides a lower bound on the variance of any unbiased estimator $\hat{\Theta}$, stating that $(\Delta \hat{\Theta})^{2}\geq 1/(\mu \mathcal{F}_C)$, where $\mu$ is the number of independent repetitions and $\mathcal{F}_C$ is the classical Fisher information (CFI). 
When optimized over measurements, the CFI can saturate the quantum Fisher information (QFI).
That is, there exists an optimal measurement for which $\mathcal{F}_C = \mathcal{F}_Q$, where the QFI is defined as  $\mathcal{F}_Q = \Tr{\rho_{\theta} L_{\theta}^{2}}$, with $L_\theta$ being the symmetric logarithmic derivative (SLD) operator~\cite{Helstrom1969}.

Hence, the ultimate estimation precision is given by the quantum Cram\' er–Rao bound (QCRB)~\cite{Holevo1982,Helstrom1969,PhysRevLett.72.3439}
\begin{eqnarray}
    (\Delta \Theta)^{2}\geq  \frac{1}{\mu \mathcal{F}_C} \geq \frac{1}{\mu \mathcal{F}_Q}. 
\end{eqnarray}
For a large number of independent repetitions $\mu$, the first inequality can be saturated by using a maximum likelihood estimator.  
Thus, achieving the optimal precision set by the QCRB requires finding an optimal measurement $\{M_x\}$ such that $\mathcal{F}_C(\rho_\theta,\{M_x\}) = \mathcal{F}_Q(\rho_\theta)$.

One approach to realizing such optimal measurements is via projective measurements onto the eigenstates of the SLD operator $L_\theta$~\cite{PhysRevLett.72.3439}. 
However, these measurements are typically highly nonlocal~\cite{e20070485}, posing significant experimental challenges.
The following theorem presents an alternative construction of optimal measurements based on the state symmetry $S$.

\textbf{Theorem~1} 
Let $\rho_0$ be a probe state with symmetry $S$ [Eq.~\eqref{State_Symmetry}], and let $\rho_\theta$ be the corresponding parameter-encoded state [Eq.~\eqref{encoded_state}].  
Suppose the encoding Hamiltonian $H$ and a set of rank-1 POVM elements $\{M_x\}_{x\in\mathcal X}$ satisfy
\begin{itemize}
\item[(i)] $ P\{S,H\} P = 0$, and
\item[(ii)]  $ P\{M_x,H\} P= 0$, for all $x\in\mathcal X$,
\end{itemize}
where $\{A,B\} = AB + BA$ and 
$P = \sum_{j: p_j \neq 0} \ket{\psi_j}\bra{\psi_j}$ is the projector onto the support of $\rho_0$.
Then the QCRB for $\rho_\theta$ can be saturated by performing the measurement
$\{U_\theta M_x U_\theta^\dagger\}_{x \in \mathcal X}$.

The proof is given in the Methods section.  
We note that restricting to rank-1 POVM elements entails no loss of generality, 
as any POVM saturating the QCRB can be replaced by an equivalent rank-1 POVM refinement~\cite{PhysRevA.111.022436}.  
We also note that condition~(i) is redundant, as it follows from condition~(ii) together with $PS=SP=P$. 
In particular, summing the condition~(ii) and using this identity immediately yields condition~(i), i.e., $\sum_{x\in\mathcal X}P\{M_x,H\}P=2PHP=P\{S,H\}P$.
Nevertheless, we keep condition~(i) because it makes explicit the symmetry constraint imposed on the Hamiltonian.
This separation is useful when applying Theorem~1 as a constructive method for identifying optimal measurement operators for parameter estimation: condition~(i) captures the symmetry constraint on the Hamiltonian, while condition~(ii) specifies the measurement compatibility requirement.

As an example, we show that the symmetry-projection theorem of Ref.~\cite{PhysRevLett.133.260402} 
is recovered as a special case of Theorem~1.
The symmetry-projection theorem states that, for a projector $\Pi$, if (a) $\Supp{\rho} \subseteq \Supp{\Pi}$ and (b) $\Pi H \Pi = 0$, 
then the QCRB can be saturated by the binary measurement  $\{\Pi,I-\Pi\}$ with $I$ being the identity operator.

Note that a projector $\Pi$ satisfying condition (a) represents a special case of the state symmetry defined in Eq.~\eqref{State_Symmetry}, highlighting the first key distinction between Theorem~1 and the symmetry-projection theorem.  
Even when the symmetry operator is restricted to a projector, i.e., $S=\Pi$, Theorem~1 remains more general.  
The symmetry-projection theorem restricts the optimal measurement to the form $M_x\in \{\Pi,I-\Pi\}$, whereas condition (ii) in Theorem~1 does not impose such a restriction.

Moreover, for the same choice of symmetry operator $S=\Pi$, and  taking the measurement to be a rank-1 POVM refinement of  $\{\Pi,I-\Pi\}$, conditions (a) and (b) satisfy constraints (i) and (ii) (see Supplementary Note~1), illustrating that the symmetry-projection theorem can be viewed as a simplified special case.   
This distinction can also be understood from the supports on which the constraints act: constraints (i) and (ii) apply on the support of $\rho_0$, whereas the symmetry-projection theorem imposes condition (b) on the support of $\Pi$.  
Because condition (a) implies $\Supp{\rho_0} \subseteq \Supp{\Pi}$, the constraints in Theorem~1 are strictly weaker than those in the symmetry-projection theorem when $\Supp{\rho_0} \subsetneq \Supp{\Pi}$.
This allows for the identification of a broader class of Hamiltonians for which a rank-1 POVM refinement of $\{\Pi,I-\Pi\}$  serves as the optimal measurement.

As an example, we consider a two-qubit probe state $\rho_0 = \ket{\phi_+}\bra{\phi_+}$ with $\ket{\phi_\pm} = (\ket{00} \pm \ket{11})/\sqrt{2}$.   
Let us take the even-parity projector $\Pi = \ket{00}\bra{00} + \ket{11}\bra{11}$ as the probe-state symmetry, i.e., $\Pi \ket{\phi_+} = \ket{\phi_+}$.   
However, the symmetry-projection theorem cannot identify the Hamiltonian $H = \ket{\phi_+}\bra{\phi_-} + \ket{\phi_-}\bra{\phi_+}$, since condition (b) is violated: $\Pi H \Pi = H \neq 0$.   
In contrast, Theorem~1 identifies this $H$ as compatible with conditions (i) and (ii), because $P\{S,H\}P =2PHP=0$ with $S=\Pi$ and $P=\ket{\phi_+}\bra{\phi_+}$, and condition (ii) can be satisfied by a rank-1 POVM refinement of $\{\Pi,I-\Pi\}$.  
A detailed verification of QCRB saturation is provided in Supplementary
Note~1.  
This explicit protocol demonstrates that Theorem~1 is more general than the symmetry-projection theorem.

It may appear that Theorem~1 applies only to unitary encoding processes~\eqref{encoded_state}.  
The following Theorem~2 demonstrates an important extension of Theorem~1 to more general parametrized pure states.

\textbf{Theorem~2}
For a general parametrized pure state $|\psi_\theta\rangle$, if it admits an expansion in a fixed basis $\{|v_j\rangle\}$ of the form
\begin{equation}\label{general psitheta}
|\psi_\theta\rangle = \sum_j \psi_j(\theta) |v_j\rangle, \quad \psi_j(\theta) \in \mathbb{R},
\end{equation}
with basis states $\ket{v_j}$ independent of $\theta$,  
then the QCRB can be saturated by the optimal projective measurement $\{M_j = |v_j\rangle\langle v_j|\}$.

The idea of the proof is as follows.  
For a fixed parameter value $\theta_0$, the parametrized state $|\psi_\theta\rangle$ can be  approximated by a unitary-encoded state $\ket{\phi_\theta} = e^{-i \delta_\theta H/2} \ket{\psi_{\theta_0}}$,
with $\delta_\theta=\theta-\theta_0$ and the effective encoding Hamiltonian being $H=2i(\ket{\dot\psi_{\theta_0}}\bra{\psi_{\theta_0}}-\ket{\psi_{\theta_0}}\bra{\dot\psi_{\theta_0}})$.
Here, $\ket{\dot\psi_{\theta_0}}=d\ket{\psi_\theta}/d\theta|_{\theta_0}$.
A key observation is that the expansion coefficients $\psi_j(\theta)$ are real. 
This guarantees the orthogonality condition $\langle \psi_{\theta_0}|\dot\psi_{\theta_0}\rangle = 0$, so that $|\psi_\theta\rangle$ coincides with $|\phi_\theta\rangle$ to first order in $\delta_\theta$.  
Since $|\phi_\theta\rangle$ corresponds to a unitary encoding process, we then verify that the conditions of Theorem~1 are satisfied for $|\phi_\theta\rangle$ in Methods, which implies that the same local projective measurement $\{M_j = |v_j\rangle \langle v_j|\}$ saturates the QCRB for $|\psi_\theta\rangle$ at $\theta = \theta_0$.
Since $\theta_0$ can be arbitrary, we prove Theorem 2.

We now apply Theorem 2 to the Hatano–Nelson model,
\begin{equation}
 H_{\mathrm{HN}}=\sum_{j=1}^N \left(J_L|j\rangle \langle j+1|-J_R|j+1\rangle \langle j|\right),
\end{equation}
a paradigmatic non-Hermitian topological system that can serve as a probe for bulk Hamiltonian parameters with quantum-enhanced sensitivity reaching HL scaling~\cite{Sarkar_2024}.
We consider the estimation of the asymmetry parameter $\theta=J_R/J_L$ with $J_R,J_L>0$.
Under open boundary conditions, the eigenstates of 
$H_{\mathrm{HN}}$ are
\begin{equation}   |\psi^{(m)}\rangle=\sum_{1\le j\le N}C_m \sqrt{\theta^j}\sin \left(\frac{jm\pi}{N+1}\right)|j\rangle,
\end{equation}
where $C_m\in\mathbb R$ is a normalization constant and $\{|j\rangle\}$ denotes the position basis.
As shown in Ref.~\cite{Sarkar_2024}, at the critical point $\theta\to\theta_c=1$, these eigenstates enable parameter estimation with HL-scaling precision.
Importantly, the wavefunction amplitudes $\psi_j(\theta)=C_m \sqrt{\theta^j} \sin(jm\pi/(N+1))$ are real-valued. 
Therefore, by Theorem 2, the optimal measurement that saturates the QCRB is given by the set of projectors onto the position basis, $M_j=\ket{j}\bra{j}$.

Having established the theory of optimal measurements, we now turn to a more practical scenario involving local measurements, which are typically far more experimentally accessible than fully nonlocal ones, as illustrated in Fig.~\ref{fig:1}.
This motivates us to develop a refined version of Theorem~1 that explicitly incorporates locality constraints on the measurement operators.

~\\
\noindent  
\textbf{Local State Symmetry-induced Optimal Local Measurements}
~\\
\noindent 
We now consider an $N$-qudit probe state $\rho_0\in\mathcal H\equiv\otimes_{j=1}^N\mathcal H_j$, where $\mathcal H_j$ denotes the local Hilbert space of the $j$-th qudit with dimension $d$.
The state $\rho_0$ is said to possess a local state symmetry if its symmetry operator $S \in \mathcal H$ [see Eq.~\eqref{State_Symmetry}] factorizes as
\begin{align}\label{local_symmetry}
S=\bigotimes_{j=1}^N S_j, \ S_j\in\mathcal H_j.    
\end{align}
Analogously, a measurement element $M_{\vec x}\in\mathcal H$ is called local if it can be expressed as
\begin{eqnarray}\label{LM}
M_{\vec x}=\bigotimes_{j=1}^N M_{x_j},\ M_{x_j}\in\mathcal H_j,    
\end{eqnarray}
where $\vec x = (x_1, \dots, x_N)$ collects the local outcomes of each qudit.
To highlight the role of local symmetries in parameter estimation, we focus on linear encoding Hamiltonians of the form $H = \sum_{j=1}^N H_j$ with $H_j \in \mathcal H_j$, and consider experimentally feasible rank-1 local projective measurements.  

\textbf{Theorem~3}
Let $\rho_0$ be an $N$-qudit state exhibiting the local symmetry~\eqref{local_symmetry}.  
For each qudit, suppose there exists a local Hamiltonian $H_j$ and $d$ different local observables $O_{x_j}$ acting on $\mathcal{H}_j$, which mutually commute and satisfy
\begin{align}\label{project-rank-1}
\Tr {O_{x_j}} = 2-d, \quad
O_{x_j}^2 = I,
\end{align}
where $I$ is the identity on $\mathcal{H}_j$.  
We define $S_{\bar{j}}=\bigotimes_{k\neq j}S_k$, $O_\alpha=\bigotimes_{j\in \alpha}O_{x_j}$ and $P=\sum_{j:p_j\neq0}|\psi_j\rangle \langle \psi_j|$ the projector onto the support of $\rho_0$. If the compatibility conditions
\begin{align}\label{local_conditions}
    &P\{S_j,H_j\}\otimes S_{\bar{j}}P=0,\nonumber\\
    &P\{H_j,O_{\alpha}\}P=0,\quad \forall\alpha\subseteq\{1,2,\ldots,N\}, \alpha\neq \emptyset.
\end{align}
are satisfied for all $j$, then, the encoded state
\begin{align}
\rho_\theta = U_\theta \rho U_\theta^\dagger, \qquad
U_\theta =\bigotimes_{j=1}^N  \exp\!\left(- i \theta  H_j/2 \right), 
\end{align}
can be estimated optimally using the optimal local measurement 
$\{U_\theta (\bigotimes_j M_{x_j}) U_\theta^\dag\}_{\vec x\in\mathcal X} $ with
\begin{align}\label{M_O}
M_{x_j} = \frac{I + O_{x_j}}{2}.
\end{align}

Theorem~3 can be viewed as the local version of Theorem~1: the compatibility conditions~\eqref{local_conditions} in Theorem~3 correspond to those in Theorem~1, specialized to local operators.
The constraints~\eqref{project-rank-1} in Theorem~3 ensure that the operators $M_{x_j}$ defined in Eq.~\eqref{M_O} are rank-1 projectors (see Supplementary
Note~2).

We now illustrate how Theorem~3 provides a simple procedure for constructing estimation protocols and optimal local measurements from local state symmetries.
In this work, we focus on $N$-qubit systems, while a discussion of qudit systems is provided in Supplementary Note~2.
For $N$-qubit systems, the rank-1 projector constraint~\eqref{project-rank-1} is naturally satisfied by choosing local observables $O_{j}\in\{\vec n_j\cdot\vec \sigma_j: |\vec n_j|=1\}$, where $\vec \sigma_j=(X_j,Y_j,Z_j)$ denotes the Pauli operators on the $j$-th qubit.
The corresponding local projective measurements can then be written as
$M_{j,\pm}=(I\pm O_j)/2$, which correspond to measurements along the directions $\pm \vec n_j$.

Writing 
\begin{equation}
S_j = \vec{s}_j \cdot \vec{\sigma}_j, \quad 
H_j = \vec{h}_j \cdot \vec{\sigma}_j, \quad 
O_j = \vec{o}_j \cdot \vec{\sigma}_j,
\end{equation}
the local compatibility conditions~\eqref{local_conditions} can be satisfied by
\begin{equation}\label{qubit}
\vec{s}_j \perp \vec{h}_j, \quad \vec{o}_j = \vec{s}_j,
\end{equation}
where $\vec{a} \perp \vec{b}$ denotes orthogonality with respect to the standard inner product.  
Indeed, the orthogonality relations $\vec{s}_j \perp \vec{h}_j$ and $\vec{o}_j \perp \vec{h}_j$ imply $\{S_j,H_j\}=0$ and $\{H_j,O_\alpha\}=0$ whenever $j\in\alpha$.
For $j\notin \alpha$, using $PS=SP=S$ and $\vec{o}_j = \vec{s}_j$, we obtain $2P\{H_j,O_\alpha\}P=P\{S,\{H_j,O_\alpha\}\}P=P\{O_\alpha,\{H_j,S\}\}P=0$ where the last equality follows from $\vec{s}_j \perp \vec{h}_j$. 
Thus, the simplified conditions~\eqref{qubit} hold for any support of the initial state $\rho_0$.
This feature becomes particularly advantageous under realistic noisy dynamics, where the support of the probe state typically expands. 
As these conditions hold for any projector $P$, they provide a robust and practical criterion for identifying optimal measurements in noisy settings, as discussed later.

Equation~\eqref{qubit} provides a constructive recipe for designing estimation protocols with optimal local measurements: given a local symmetry $\bigotimes_jS_j$ of the probe state, one may choose a local encoding Hamiltonian $H_j$ orthogonal to $S_j$, and subsequently choose a local measurement observable $O_j$ to be equal to $S_j$.
As an illustrative example, we consider  the GHZ state $ \rho_0 = |\mathrm{GHZ}\rangle\langle \mathrm{GHZ}|$, where $\ket{\rm GHZ}=(\ket{0}^{\otimes N}+\ket{1}^{\otimes N})/\sqrt{2}$, with $Z_j \ket{0 }_j = \ket{0}_j$ and $Z_j \ket{1 }_j = -\ket{1}_j$. 
The operator $S=\bigotimes_{j=1}^N X_j$ clearly defines a local state symmetry for the GHZ state. 
Applying the simplified compatibility conditions~\eqref{qubit}, we may choose a local encoding Hamiltonian
$ H = \sum_{j=1}^N Z_j $ and local measurement observables $O_j=X_j$, thereby saturating the QCRB with local measurements.

Note that $S = \bigotimes_{j=1}^N X_j$ is a stabilizer of the GHZ state. 
This example shows that, for any stabilizer state, one can directly use its stabilizers as local state symmetries to construct optimal local estimation protocols by using Eq.~\eqref{qubit}.  
Next, we consider graph states to illustrate how stabilizer theory can be used to design estimation protocols that achieve HL scaling precision by using optimal local measurements.

~\\
\noindent \textbf{Heisenberg-Limited Estimation with Graph States under Optimal Local Measurements}

\noindent
Let $\mathbb G=(\mathcal N,\mathcal E)$ denote a connected graph with vertex set $\mathcal N = \{1,2,\dots,N\}$ and edge set $\mathcal E$, where each vertex $j \in \mathcal N$ represents a qubit, and each edge $(i,j) \in \mathcal E$ indicates a connection between qubits $i$ and $j$.  
For each vertex $j \in \mathcal N$, the corresponding stabilizer generator is defined as
\begin{equation}
K_j = X_j \bigotimes_{i \in \mathcal A_j} Z_i,
\end{equation}
where $\mathcal A_j \subset \mathcal N$ denotes the set of vertices adjacent to $j$. 
An $N$-qubit graph state $|\mathbb G\rangle$ associated with the graph $\mathbb G$ is defined as the common $+1$ eigenstate of all stabilizer generators, i.e., 
\begin{align}
    K_j \ket{\mathbb G} = \ket{\mathbb G}, \quad \forall j \in \mathcal N.
\end{align}

\begin{figure}[t]
\begin{centering}
\includegraphics[scale=0.19]{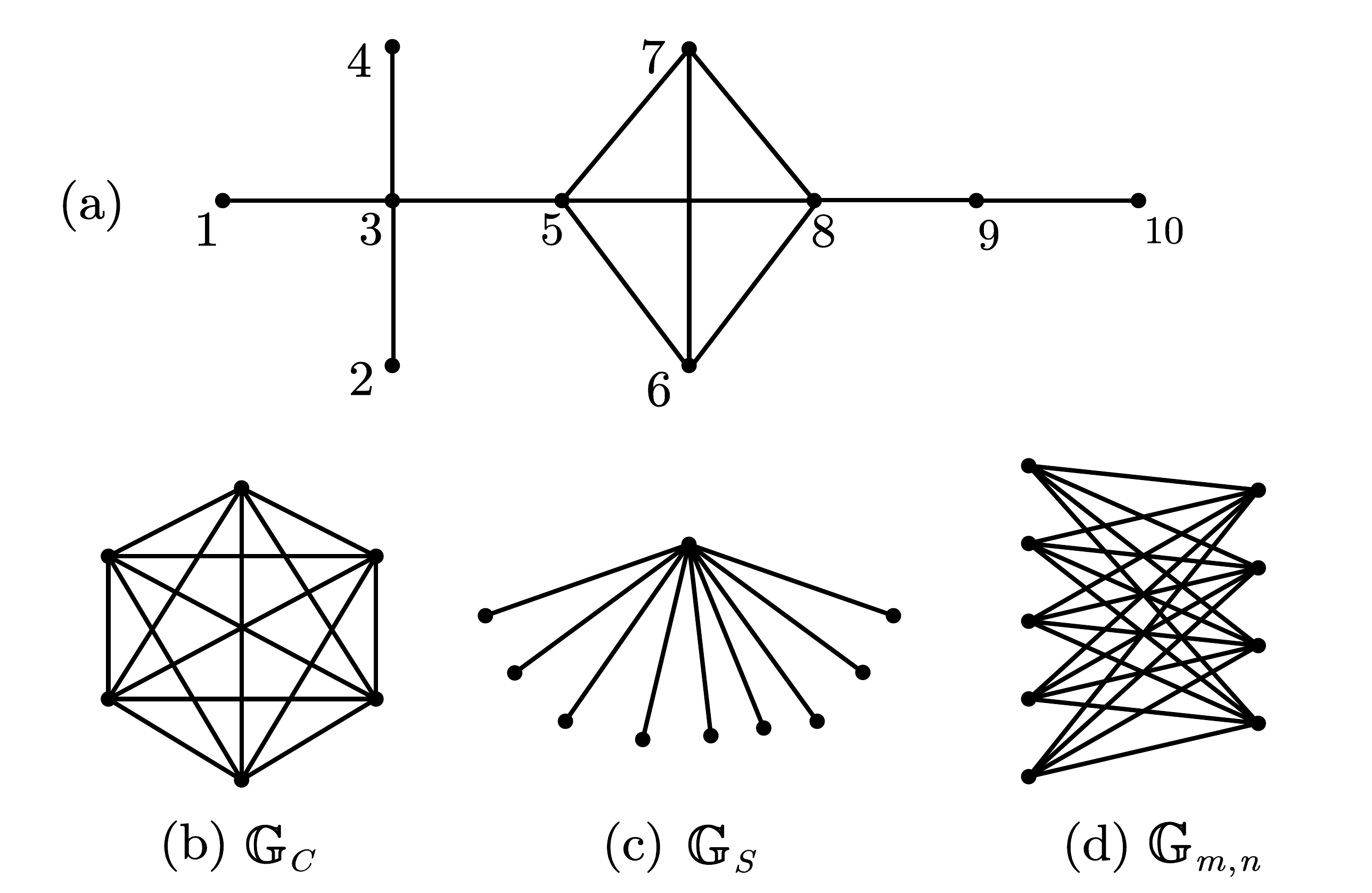}
\par\end{centering}
\caption{\label{fig2} 
Examples of graphs used in this work: (a) an example graph with highlighted vertex types, (b) the complete graph $\mathbb G_C$, (c) the star graph $\mathbb G_S$, and (d) the complete bipartite graph  $\mathbb G_{m,n}$.
%
}
\end{figure}

Exploiting this property, for each nonempty subset of vertices $\alpha \subseteq \mathcal N$, one can define a state-symmetry operator $S(\alpha)$ as the product of the stabilizer generators corresponding to $\alpha$,
\begin{align}\label{S_alpha_1}
S(\alpha) = \prod_{j \in \alpha\subseteq \mathcal N} K_j,
\end{align}
which satisfies $S(\alpha)\ket{\mathbb G} = \ket{\mathbb G}$.
Using the algebraic relations of the Pauli operators, $S(\alpha)$ can be expressed in a local form (up to a global phase) as in Eq.~\eqref{local_symmetry}
\begin{align}\label{local_symmetry_graph}
S(\alpha) \propto X_{\beta_X} I_{\beta_I} Y_{\beta_Y} Z_{\beta_Z},
\end{align}
where $\{\beta_X,\beta_I,\beta_Y,\beta_Z\}$ form a partition of the qubit set $\mathcal N$ determined by the choice of $\alpha$.
The shorthand notation $O_{\beta_O}$ denotes the tensor product of $O$ over all qubits in $\beta_O$, i.e.,
$O_{\beta_O}=\bigotimes_{j\in\beta_O} O_j$ with $O\in{X,Y,Z,I}$.
For example, taking $\alpha = \{1,2\}$ in Fig.~\ref{fig2}(a), we have $S(\alpha) = K_1 K_2 = (X_1 Z_3)(X_2 Z_3) \propto X_1 X_2$,
so that $\beta_X = \{1,2\}$.

Given the local state symmetry~\eqref{local_symmetry_graph}, the compatibility conditions~\eqref{qubit} can be used to construct the encoding Hamiltonian and the corresponding optimal local measurements.  
We emphasize that the orthogonality condition between the local symmetry operator $S_j$ and the local Hamiltonian $H_j$ indicates that the such estimation protocols can be realized by many different Hamiltonians.  
Here, we focus on the special Hamiltonian that maximizes the QFI among all protocols satisfying the compatibility conditions~\eqref{qubit}.  
A detailed discussion is provided in the  Supplementary
Note~3.

{\bf Protocol 1}
For an $N$-qubit ($N\ge 3$) connected graph state $|\mathbb G\rangle$ and any nonempty subset $\alpha \subseteq \mathcal N$, the compatibility conditions~\eqref{qubit} define an estimation protocol for $\rho_\theta = e^{-i\frac{\theta}{2}H(\alpha)} |\mathbb G\rangle \langle \mathbb G| e^{i\frac{\theta}{2}H(\alpha)}
$ with the encoding Hamiltonian
\begin{align}\label{Halpha zyx}
H(\alpha)\!=\! 
\sum_{j \in (\beta_X \cup \beta_Y) \cap \mathcal R}\!Z_j +
\sum_{\substack{j \in (\beta_X \backslash   \mathcal R) \\
                  \cup (\beta_Z \backslash  \mathcal T_{0})}}
\!Y_j +
\sum_{\substack{j \in (\beta_Y \backslash   \mathcal R) \\
                  \cup (\beta_Z \cap \mathcal T_{0})}}
\!X_j, 
\end{align}
where the sets $\beta_j$ ($j=X,Y,Z$) are determined by the local state symmetry in Eqs.~\eqref{S_alpha_1} and \eqref{local_symmetry_graph}, with $\beta_j \backslash \mathcal R$ denoting vertices in $\beta_j$ outside the root vertices $\mathcal R$.  
Here, $\mathcal R$ and $\mathcal T_0$ denote root vertices and vertices without true twins, respectively. 
The optimal local measurements can be determined using the compatibility conditions~\eqref{qubit}. 
The corresponding QFI is derived in Supplementary Note~3:
\begin{align}\label{Fmax1HK}
&\mathcal{F}_Q(\mathbb G,\alpha) = \sum_{j=X,Y,Z}|\beta_j| +\sum_m f\left(|\mathcal T_{f;m} \cap (\beta_Y\cup\beta_Z)|\right)\n
&+ \sum_m f\left(|\mathcal T_{t;m} \cap (\beta_X\cup\beta_Z)|\right)+2|\mathcal L\cap (\beta_Y\cup\beta_Z)|,
\end{align}
with $f(x)=x(x-1)$ reproducing the $x^2$ scaling for $x\gg 1$ and $|\gamma|$ denoting the cardinality of set $\gamma$. 
The sets $\{\mathcal T_{f;m}\}_m$ and $\{\mathcal T_{t;m}\}_m$ partition $\mathcal N=\{1,\dots,N\}$ according to false- and true-twin relations, respectively.

Before analyzing Protocol~1, we introduce several key graph-theoretic notions using the example in Fig.~\ref{fig2}(a).
\\
$\blacksquare$ \textbf{Leaf vertex.} A vertex connected to exactly one other vertex.
The set of all leaf vertices is denoted by $\mathcal L$, e.g.,  $\mathcal L=\{1,2, 4,10\}$ in Fig.~\ref{fig2}(a).
\\
$\blacksquare$ \textbf{Root vertex.} A non-leaf vertex adjacent to at least one leaf vertex. The set of root vertices is denoted by $\mathcal R$, e.g.,  $\mathcal R=\{3,9\}$ in Fig.~\ref{fig2}(a).
\\
$\blacksquare$ \textbf{True and false twins.} Two vertices are true twins if they share the same neighbors including each other, and false twins if they share the same neighbors excluding each other.
For example, in Fig. \ref{fig2}(a), vertices 6 and 7 form a pair of true twins, whereas vertices 1 and 2 form a pair of false twins.
The partitions of true and false twins are denoted by $\{\mathcal T_{t;m}\}_m$ and $\{\mathcal T_{f;m}\}_m$. 
Then, the set $\mathcal T_0$ of vertices without true twins can be expressed as $\mathcal T_0=\bigcup_{m:|\mathcal T_{t;m}|=1} \mathcal T_{t;m}$.

As an illustration of Protocol~1, we consider the ten-qubit graph state shown in Fig.~\ref{fig2}(a), for  which  $\mathcal R=\{3,9\}$, $\mathcal L=\{1,2,4,10\}$, and $\mathcal T_0=\{1,2,3,4,5,8,9,10\}$.

The partitions with respect to the true- and false-twin relations are given by
$\mathcal T_{t}=\{1|2|3|4|5|67|8|9|10\}$
and
$\mathcal T_{f}=\{124|3|5|6|7|8|9|10\}$,
respectively.
Here we adopt a compact notation for the partitions, where vertices listed within the same block (without delimiters) belong to the same equivalence class.
Choosing the subset $\alpha=\{3,6,9,10\}$, the corresponding local state symmetry {$\mathcal{S}(\alpha)\propto X_{\{3,6\}}I_{\{5,8\}}Y_{\{9,10\}}Z_{\{1,2,4,7\}}$} is obtained from definition~\eqref{S_alpha_1}, which identifies
$\beta_X=\{3,6\}$,
$\beta_I=\{5,8\}$,
$\beta_Y=\{9,10\}$,
and
$\beta_Z=\{1,2,4,7\}$
according to Eq.~\eqref{local_symmetry_graph}.
Then, by Eq.~\eqref{Halpha zyx}, the optimal local protocol generated by the Hamiltonian
$H=Z_3+Z_9+Y_6+Y_7+X_1+X_2+X_4+X_{10}$
yields QFI
$\mathcal{F}_Q=24$.
The QCRB can be saturated by performing local $X$- measurement on qubits $\{3,6\}$, local $Y$- measurement on qubits $\{9,10\}$,
$Z$-measurement on qubits $\{1,2,4,7\}$,
and arbitrary measurements on qubits $\{5,8\}$,
as dictated by the compatibility condition~\eqref{qubit}.

Several observations follow from Eqs.~\eqref{Halpha zyx} and \eqref{Fmax1HK}.
%
%
First, Eqs.~\eqref{Halpha zyx} and \eqref{Fmax1HK} show that qubits belonging to the subset $\beta_I$ contribute trivially to the QFI, since the corresponding state-symmetry operator $S(\alpha)$ in Eq.~\eqref{local_symmetry_graph} acts as the identity on these qubits.
To enhance the QFI, the subset $\alpha$ should therefore be chosen such that $S(\alpha)$ acts nontrivially on all qubits.
Second, the function $f(x)=x(x-1)$ appearing in Eq.~\eqref{Fmax1HK} explicitly depends on the sizes of the partitions  $\{\mathcal T_{f;m}\}_m$ and $\{\mathcal T_{t;m}\}_m$ defined by the false- and true-twin relations.
Consequently, twin structures play a crucial role in achieving Heisenberg-limited scaling, $\mathcal F_Q \sim N^2$.
This scaling can be realized by appropriately matching the twin partitions $\{\mathcal T_{f;m}\}_m$ and $\{\mathcal T_{t;m}\}_m$ with the subsets $\beta_j$ determined by the state-symmetry operator $S(\alpha)$, see Eqs.~\eqref{local_symmetry_graph} and \eqref{Fmax1HK}.

Explicitly, the QFI in Eq.~\eqref{Fmax1HK} is upper bounded as
\begin{align}\label{Fmax2GN}
 \mathcal{F}_Q(\alpha,\mathbb G) \!\leq\!  N +\sum_m f\left(|\mathcal T_{f;m} |\right)\!+\! \sum_m f\left(|\mathcal T_{t;m}|\right)
\!+\! 2|\mathcal L|,
\end{align}
which depends only on the graph structure of $\mathbb G$.
This upper bound is saturated if there exists a state-symmetry operator
$S(\alpha)\propto X_{\beta_X} I_{\beta_I} Y_{\beta_Y} Z_{\beta_Z}$
such that
\begin{itemize}
\item[(i)] $\mathcal L \subseteq\beta_Y\cup\beta_Z$,\quad 
(ii) $ \beta_I=\emptyset$, 
\item[(iii)]  $ \mathcal T_{f;m}\subseteq\beta_Y\cup\beta_Z$ for all $m$, and
\item[(iv)]  $ \mathcal T_{t;m}\subseteq\beta_X\cup\beta_Z$ for all $m$.
\end{itemize}
Conditions~(i) and (ii) ensures that $S(\alpha)$ acts nontrivially on all qubits and that all leaf vertices belong to $\beta_Y \cup \beta_Z$.
Conditions~(iii) and~(iv) require that all false twins and true twins  belong to $\beta_Y \cup \beta_Z$ and $\beta_X \cup \beta_Z$, respectively.
According to Eq.~\eqref{Fmax1HK}, the contribution from leaf vertices scales at most linearly with the total number of qubits $N$. 
Thus, conditions~(iii--iv) provide a practical guideline for identifying graph structures that achieve HL scaling.

Furthermore, Eq.~\eqref{Fmax2GN} also clarifies the necessary condition for HL scaling.
If all twin sets remain subextensive, i.e., $|\mathcal T_{f;m}| = o(N)$ and 
$|\mathcal T_{t;m}| = o(N)$ for all $m$, then the right-hand side of 
Eq.~\eqref{Fmax2GN} remains subquadratic, i.e., 
$\mathcal{F}_Q = o(N^2)$.
Therefore, HL scaling $\mathcal{F}_Q\sim N^2$ can occur if and only if 
the graph contains an extensive set of mutually false twins or mutually 
true twins, i.e., $|\mathcal T_{f;m}|\sim N$ or $|\mathcal T_{t;m}|\sim N$ 
for some $m$.
The sufficiency follows directly from Eq.~\eqref{Fmax1HK}.

Using the true-twin condition~(iv), we consider the complete graph $\mathbb G_C$ on $N$ vertices, with edges between every pair of vertices, see Fig.~\ref{fig2}(b). 
All vertices in $\mathbb G_C$ are true twins, i.e., $\mathcal T_{t;1} = \mathcal N \equiv \{1,2,\dots,N\}$. 
A corresponding state-symmetry operator can be chosen from any stabilizer generator; for example, taking the first qubit gives
$S(\{1\}) = K_1 = X_1 Z_{\{2,3,\dots,N\}}$,
which yields $\beta_X = \{1\}$ and $\beta_Z = \{2,3,\dots,N\}$.
Clearly, condition~(iv) is satisfied since $\beta_X \cup \beta_Z = \{1,2,\dots,N\}$. 
From the term $\sum_m f(|\mathcal T_{t;m}|)$ in the QFI expression~\eqref{Fmax1HK}, we then deduce that $\mathcal F_Q \sim f(|\mathcal T_{t;1}|) \sim N^2$,
indicating that the HL scaling is achieved.

By using the false-twin condition~(iii), we consider the star graph $\mathbb G_S$, consisting of a central vertex, denoted by $1$, connected to $N-1$ peripheral vertices, with no edges among the peripheral vertices, see Fig.~\ref{fig2}(c). 
In this graph, all $N-1$ peripheral vertices are false twins, giving the partition $\mathcal T_{f;1} = \{2,3,\dots,N\}$ and $\mathcal T_{f;2} = \{1\}$ for the central vertex. 
We choose the state-symmetry operator from the stabilizer generator associated with the central vertex, i.e., $S(\{1\}) = X_1 Z_{\{2,3,\dots,N\}}$, which yields $\beta_Z = \{2,3,\dots,N\}$.
Since $\mathcal T_{f;1} \subseteq \beta_Z$, Eq.~\eqref{Fmax1HK} implies that $\mathcal F_Q \sim f(|\mathcal T_{f;1}|) \sim N^2$, with the quadratic contribution arising from the $N-1$ false twins.

Another example achieving HL scaling is the complete bipartite graph $\mathbb G_{m,N-m}$, where the $N$ vertices are divided into two disjoint sets: the first set with $m$ vertices labeled $1,\dots,m$ and the second set with $N-m$ vertices labeled $m+1,\dots,N$. 
Every vertex in the first set is connected to every vertex in the second set, with no edges within each set, see Fig.~\ref{fig2}(d). 
Note that the star graph corresponds to the special case $m=1$.
In $\mathbb G_{m,N-m}$, the two disjoint sets form the false-twin partitions $\mathcal T_{f;1} = \{1,2,\dots,m\}, \quad 
\mathcal T_{f;2} = \{m+1,m+2,\dots,N\}$.
We consider a state-symmetry operator constructed as the product of stabilizer generators from each subset, e.g., $S(\{1,m+1\}) \propto Y_{\{1,m+1\}} Z_{\{2,\dots,m,m+2,\dots,N\}}$.
It follows that $\mathcal T_{f;1}   \subseteq \beta_Y \cup \beta_Z$, $\mathcal T_{f;2}   \subseteq \beta_Y \cup \beta_Z$, and thus  Eq.~\eqref{Fmax1HK} implies $\mathcal F_Q \sim f(|\mathcal T_{f;1}|) + f(|\mathcal T_{f;2}|) \sim N^2$,
demonstrating that the complete bipartite graph also achieves HL scaling.

From the above examples, it is clear that twin structures in the graph play a crucial role in achieving HL scaling.
Motivated by this, we propose two graph-connection rules that enable the recursive combination of smaller graphs into larger ones, thereby providing a scalable construction that preserves the HL scaling.

The first one is the weak connection rule. 
Let $\mathbb G$ and $\mathbb G'$ be two graphs with vertex sets $\mathcal N$ and $\mathcal N'$, respectively. 
We select $m$ vertices from each graph and connect them one-to-one, generating a new graph 
$\mathbb G_{w}=\mathbb G\oplus_w\mathbb G'$.
Here, $\oplus_w$ denotes the weak connection rule, meaning that $m \ll |\mathcal N|$ and $m \ll |\mathcal N'|$. 
Since only a few vertices are connected, the neighborhoods of most vertices remain unchanged, and the twin structures of $\mathbb G_w$ can be approximated by the disjoint union of those in $\mathbb G$ and $\mathbb G'$ when $|\mathcal N|,|\mathcal N'|\gg1$. 
Moreover, the product of the state-symmetry operators is approximately preserved under the weak connection rule, i.e.,
$S(\alpha,\mathbb G_w)S(\alpha',\mathbb G_w)\simeq S(\alpha,\mathbb G)S(\alpha',\mathbb G')$,  where $S(\alpha,\mathbb G)$ denotes the state-symmetry operator determined via Eq.~\eqref{S_alpha_1} for graph $\mathbb G$.
As a consequence, the QFI approximately satisfies an additive property at the scaling level,
\begin{align}
\mathcal F_Q(\alpha\cup \alpha', \mathbb G\oplus_w\mathbb G') \sim \mathcal F_Q(\alpha, \mathbb G)+ \mathcal F_Q(\alpha', \mathbb G').
\end{align}
This implies that if $\mathbb G$ and $\mathbb G'$ individually realize HL scaling, then their weakly connected graph $\mathbb G_w$ also achieves HL scaling. 
We denote the graph states constructed in this way as W-type states. 

\begin{figure}[t]
\begin{centering}
\includegraphics[scale=0.14]{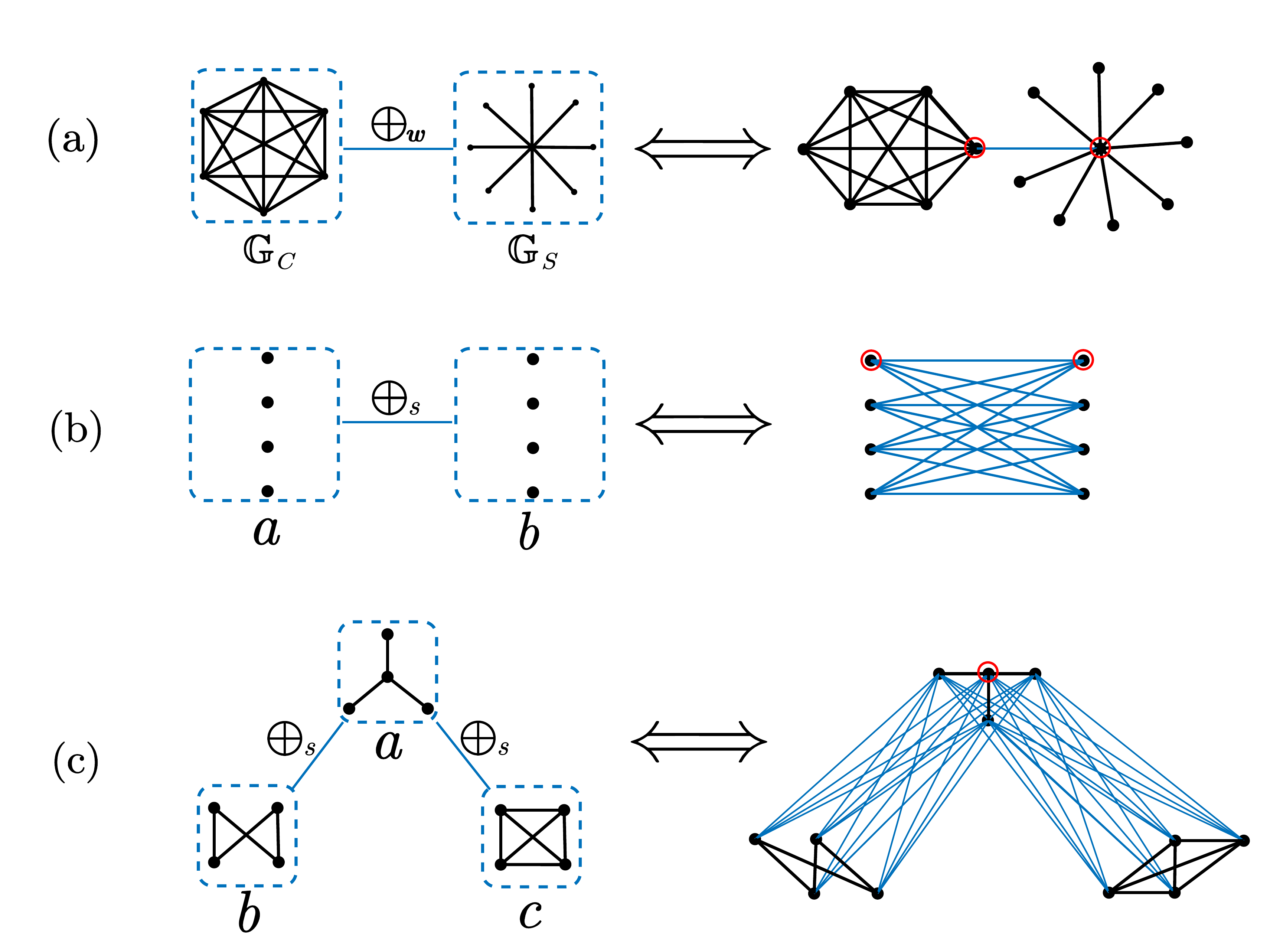}
\par\end{centering}
\caption{\label{fig3}
(a) A W-type state obtained by weakly connecting a complete graph and a star graph;  (b) A S-type state obtained from a two-vertex parent graph $a\!\leftrightarrow\! b$ with both vertices being replaced by edgeless subgraphs;
(c) A S-type state obtained from a three-vertex parent graph $b\!\leftrightarrow\! a\!\leftrightarrow\! c$, with vertices replaced by complete bipartite, star, and complete subgraphs, respectively. 
The vertices belonging to the subset $\alpha$ that determines the state symmetry~\eqref{S_alpha_1} are marked by red circles.}
\end{figure}

For instance, we consider a complete graph $\mathbb G_C$ on $N/2$ vertices 
and a star graph $\mathbb G_S$ on $N/2$ vertices.
By weakly connecting the two graphs [see Fig.~\ref{fig3}(a)], the resulting W-type state exhibits HL scaling, with the QFI calculated from Eq.~\eqref{Fmax1HK} given by $\mathcal F_Q(\alpha\cup \alpha',\mathbb G_C \oplus_w \mathbb G_S)=N^2/2-N+2$, where the vertices in $\alpha$ and $\alpha'$ are marked in red in Fig.~\ref{fig3}(a).
Since $\mathcal F(\alpha,\mathbb G_C)=\mathcal F(\alpha,\mathbb G_S)=N^2/4$, this example demonstrates the approximate additivity of the QFI under weak connections.

The second connection scheme, referred to as the strong connection rule, generates a new graph 
$\mathbb G_s = \mathbb G \oplus_s \mathbb G'$ by connecting every vertex in $\mathbb G$ to every vertex in $\mathbb G'$.
Under this rule, if vertices $i,j \in \mathcal N$ are true or false twins in $\mathbb G$, they remain true or false twins in $\mathbb G_s$.  
However, the product of the state-symmetry operators is generally not preserved under the strong connection rule, i.e., 
$S(\alpha, \mathbb G_s) S(\alpha', \mathbb G_s) \not\simeq S(\alpha, \mathbb G) S(\alpha', \mathbb G')$, since the neighborhood of each vertex $j \in \mathcal N$ in $\mathbb G$ is enlarged to further include all vertices in $\mathbb G'$ after the strong connection.
Consequently, the approximate additivity of QFI may fail, and a more restrictive selection rule is required to determine which graphs can be strongly connected while maintaining HL scaling.

In Supplementary Note~4, we establish a selection rule that guarantees HL scaling under the strong connection rule.
We begin with a connected parent graph $\mathbb G_P$ whose state-symmetry operator has the form 
$S \propto X_{\beta_X} Y_{\beta_Y} Z_{\beta_Z}$, 
acting non-trivially on every vertex. 
Accordingly, the sets ${\beta_X,\beta_Y,\beta_Z}$ form a partition of the vertex set of $\mathbb{G}_P$.
Next, each vertex $j$ of the parent graph is replaced by a subgraph according to its Pauli label:
\begin{itemize}
\item vertices in $\beta_X$ by $\mathbb G_C$, $\mathbb G_S$, or $\mathbb G_{m,n}$;
\item vertices in $\beta_Y$ by $\mathbb G_S$, $\mathbb G_{m,n}$, or $\mathbb G_E$;
\item vertices in $\beta_Z$ by $\mathbb G_C$, $\mathbb G_S$, $\mathbb G_{m,n}$, or $\mathbb G_E$.
\end{itemize}
Here, $\mathbb G_C$, $\mathbb G_S$, $\mathbb G_{m,n}$, and $\mathbb G_E$ denote the complete, star, complete bipartite, and edgeless graphs, respectively.
After substitution, every edge of the parent graph is lifted to a strong connection between the corresponding subgraphs. 
Graph states generated in this way are referred to as $S$-type states.
We require that the number of vertices in each subgraph scales linearly with the total number of qubits $N$.  
Under this condition, HL scaling of the QFI is preserved, and an estimation protocol with optimal local measurements exists.

As an example,  we consider the parent graph $\mathbb G_P$ consisting of two connected vertices $a$ and $b$, i.e., $a\!\leftrightarrow\! b$, as shown in Fig.~\ref{fig3}(b).
We choose the state-symmetry operator $S(\{a,b\}) = K_a K_b \propto Y_a Y_b$, giving $\beta_Y = \{a,b\}$.
According to the selection rule, both vertices $a$ and $b$ can be replaced by edgeless subgraphs.
Applying the strong connection rule then produces the bundled graph state shown in Fig.~\ref{fig3}(b), which also corresponds to a complete bipartition graph state and exhibits HL scaling as discussed above.
Furthermore, by applying the strong connection rule exclusively to edgeless subgraphs, one can generate other bundled graph states, as also discussed in Ref.~\cite{PhysRevLett.124.110502}.

As another example of S-type states, we consider the parent graph $\mathbb{G}_P$ consisting of three vertices $a$, $b$, and $c$, with connections $b \!\leftrightarrow\! a \!\leftrightarrow\! c$, see Fig.~\eqref{fig3}(c). 
We choose the state-symmetry operator $S(\{a\}) = X_a Z_b Z_c$, giving $\beta_X = \{a\}$ and $\beta_Z = \{b,c\}$. 
According to the selection rule, the vertices $a$, $b$, and $c$ can be replaced by a star subgraph, a complete bipartite subgraph, and a complete subgraph, respectively. 
Assuming each subgraph contains $N/3$ vertices and $m=n=N/6$ in the complete bipartite graph, the QFI of the resulting S-type state from Eq.~\eqref{Fmax1HK} exhibits HL scaling: $\mathcal{F}_Q(\alpha,\mathbb{G}_{m,n}\oplus_s\mathbb{G}_{S}\oplus_s\mathbb{G}_{C})=5N^2/18-2N/3+2$, where the vertices in $\alpha$ are marked in red in Fig.~\ref{fig3}(c).

\begin{figure}[t]
\begin{centering}
\includegraphics[scale=0.22]{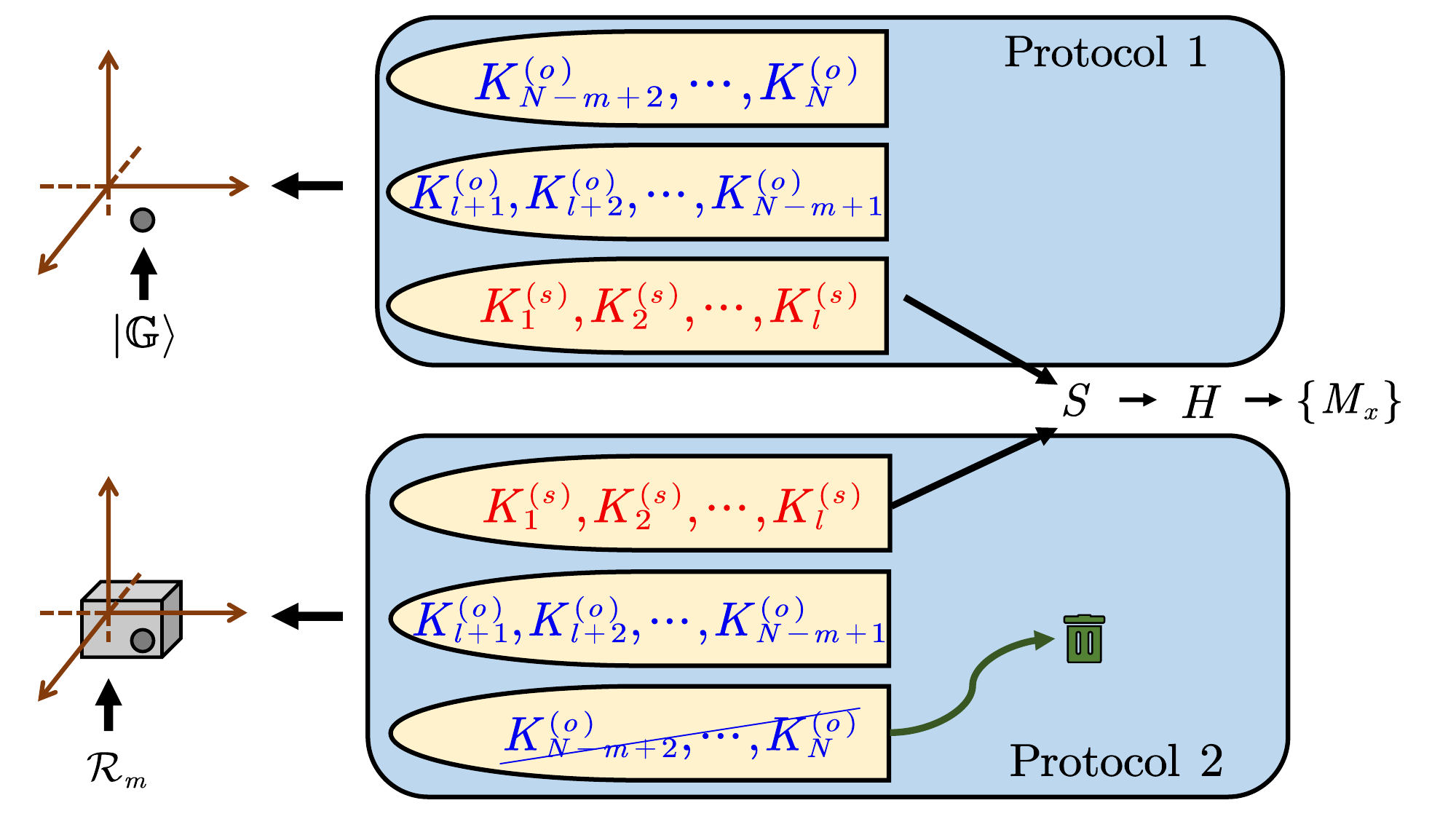}
\par\end{centering}
\caption{\label{protocol 12}
Comparison of Protocols 1 and 2.
Both protocols share the same state symmetry $S$ guaranteed by the stabilizer generators $\{K_j^{(s)}\}$. 
Accordingly, by Theorem~3 [or Eq.~\eqref{qubit}], the same Hamiltonian and local measurements ${M_x}$ can be used to construct the estimation protocol and to saturate the QCRB.
Protocol 2 differs from Protocol 1 in that it removes some of stabilizer generators irrelevant to the state symmetry, $K_{N-m+2}^{(o)},\cdots,K_{N}^{(o)}$, thereby enlarging the probe space from a single graph state $|\mathbb{G}\rangle$ to a relaxed-stabilizer subspace $\mathcal{R}_m$. 
Here, the superscripts 
$(s)$ and $(o)$ denote stabilizer generators that are respectively related and unrelated to the state-symmetry operator $S$.}
\end{figure}

\noindent
\textbf{Noise-Resilient States in the Relaxed-Stabilizer Subspace}

\noindent The results presented above indicate that optimal local measurements do not rely on the full stabilizer structure of a given graph state.
Instead, the optimal local measurements is determined by a subset of stabilizer generators that induce the local state symmetry underlying its construction.
This observation motivates a general construction strategy, see Fig~\ref{protocol 12}: relaxing stabilizer constraints while retaining the generators responsible for the relevant local state symmetry promotes a single graph state to a higher-dimensional subspace without affecting the existence of  optimal local measurements.
We refer to the resulting space as the relaxed-stabilizer subspace.

This construction has several immediate consequences.
First, relaxing stabilizer constraints increases the dimensionality of the admissible probe space, introducing additional degrees of freedom beyond those available in the original graph state.
Second, the original stabilizer state appears as a special point within the relaxed-stabilizer subspace, implying that this construction strictly generalizes the above stabilizer-based metrological protocols.
Finally, as we demonstrate below, certain states within the relaxed-stabilizer subspace exhibit enhanced robustness against noise and can surpass the original graph state.

To make the above strategy explicit, we present a representative construction based on the GHZ state, which is locally unitarily equivalent to the star graph state. 
The $N$-qubit GHZ state $\ket{ {\rm GHZ}}_N=(\ket{00\cdots 0}+\ket{11\cdots 1})/\sqrt{2}$
is stabilized by the $N$ stabilizer generators
\begin{align}\label{GHZ-stablizer}
&X_1 X_2 \cdots X_N, \n
&Z_1 Z_i, \quad i = 2, \dots, N .
\end{align}

We now construct the relaxed-stabilizer subspace $\mathcal R_m$ by partially relaxing $(m-1)$ GHZ stabilizer constraints.
To this end, we partition the $N$ qubits into $m$ disjoint subsets
\begin{align}
\mathcal{N}_1 &= \{1, \dots, N_1\}, \n
\mathcal{N}_2 &= \{N_1+1, \dots, N_1+N_2\}, \n
&\;\;\vdots \nonumber \\
\mathcal{N}_m &= \{N-N_m+1, \dots, N\},
\end{align}
with $\sum_{j=1}^m N_j = N$.
We use the notation $X_{i;j}$ to denote the Pauli-$X$ operator acting on the $j$-th qubit within the subset $\mathcal{N}_i$, and analogously for $Y_{i;j}$ and $Z_{i;j}$, e.g., $X_{2;1}=X_{N_1+1}$.
On this partition, we retain the global stabilizer
\begin{align}\label{XXX}
 X_1 X_2 \cdots X_N ,
\end{align}
and partially relax the remaining GHZ constraints by keeping only the intra-subset $Z$-type stabilizers
\begin{align}\label{ZZ}
Z_{j;1} Z_{j;2},\;
Z_{j;1} Z_{j;3},\;
\dots,\;
Z_{j;1} Z_{j;N_j},
\quad j = 1, \dots, m .
\end{align}
The total number of remaining stabilizer generators is therefore $N-(m-1)$.

Compared to the full GHZ stabilizer set~\eqref{GHZ-stablizer}, the two-body $Z$-type constraints~\eqref{ZZ} are removed across different subsets $\mathcal{N}_j$, resulting in an enlarged common $+1$ eigenspace.
We denote this space as the relaxed-stabilizer subspace $\mathcal{R}_m$.
While $\mathcal{R}_m$ spans a higher-dimensional Hilbert subspace, it preserves the same local-state symmetry $X_1 X_2 \cdots X_N$, implying that all states within the subspace support the same estimation protocol as Protocol~1, with Hamiltonian $H=\sum_{j=1}^N Z_j$ and the identical optimal local measurement  $X_1X_2\cdots X_N$.

{\bf Protocol 2}
Consider the relaxed-stabilizer subspace $\mathcal{R}_m$, defined as the common $+1$ eigenspace of the stabilizer generators in Eqs.~\eqref{XXX} and~\eqref{ZZ}.  
This subspace is spanned by the set of states
\begin{align}\label{vect}
&\ket{\vec t} = X_{\mathcal{N}_1}^{t_1} X_{\mathcal{N}_2}^{t_2} \cdots X_{\mathcal{N}_m}^{t_m} \ket{\mathrm{GHZ}}_N, 
\quad t_1 = 0, \; t_{j\neq 1} \in \{0,1\}\n 
& \vec t = (t_1,\dots,t_m) 
\end{align}
where $X_{\mathcal{N}_j}^{t_j}$ denotes the operator that applies a Pauli-$X$ to every qubit in the subset $\mathcal{N}_j$ if $t_j = 1$, and acts as the identity if $t_j = 0$.
For a general mixed state $\rho_{\rm rss}$, referred to as a relaxed-stabilizer state, supported entirely on $\mathcal{R}_m$, we can write
\begin{align}\label{state_pro_2}
\rho^{\rm rss} = \sum_{\vec t, \vec t' : t_1 = t_1' = 0} \rho_{\vec t, \vec t'} \, \ket{\vec t}\bra{\vec t'}, 
\qquad \rho^{\rm rss}_{\vec t, \vec t'} = \ex{ \vec t | \rho_{\rm rss} | \vec t' }.
\end{align}
The optimal local estimation protocol can then be realized by the unitary phase-encoding model $\rho(\theta)=e^{-i\frac{\theta}{2}\sum_jZ_j}\rho e^{i\frac{\theta}{2}\sum_j Z_j}$ 
which achieves the corresponding QFI
\begin{align}\label{QFI_subspace}
\mathcal{F}_Q(\rho^{\rm rss})=\sum_{\vec t:t_1=0}
\rho^{\rm rss}_{\vec t,\vec t}\left(N-2\sum_{j=1}^m t_j N_j\right)^{2}.
\end{align}
The derivation is provided in the Methods.

We now turn to an explicit analysis of the QFI for
relaxed-stabilizer states supported on the subspace $\mathcal R_m$.
We introduce the relative block size $\xi_j=N_j/N$, defined as the ratio of the
number of qubits in the $j$-th subset to the total number of qubits.
By defining
\begin{align}
r_{\min}\equiv \min_{\vec t:t_1=0} \left(1-2\sum_{j=1}^mt_j \xi_j  \right)^{2}, \quad \xi_j=\frac{N_j}{N},
\end{align}
Eq.~\eqref{QFI_subspace} can be lower bounded as
\begin{align}\label{QFI_subspace_lb}
\mathcal F_Q(\rho^{\rm rss})\geq r_{\min} N^2,   
\end{align}
where we have used $\sum_{\vec t:\, t_1=0}\rho_{\vec t,\vec t}=1$.
Equation~\eqref{QFI_subspace_lb} thus directly establishes the worst-case QFI determined solely by the partition structure of the
relaxed-stabilizer subspace $\mathcal R_m$.

HL scaling is therefore preserved whenever the partition defining
$\mathcal R_m$ yields a finite, $N$-independent, and nonvanishing minimal ratio $r_{\min}$.
This condition is satisfied by several natural partition families, which we now illustrate.
\\(a) Biased bipartition. For a partition with $\xi_1=(1+\delta)/2$ and $\delta>0$, one finds
$r_{\min}=\delta^2$, leading to the bound
\begin{align}\label{bias}
\mathcal F_Q(\rho^{\rm rss}) \ge \delta^2 N^2.    
\end{align}
(b) Uniform partition. For an odd number $m$ of equal-size partitions, $\xi_j=1/m$, one obtains
$r_{\min}=1/m^2$, and hence
\begin{align}\label{uniform}
\mathcal F_Q(\rho^{\rm rss}) \ge N^2/m^2.    
\end{align}
These partition families serve as benchmarks for assessing noise robustness in the following.

Next, we investigate the behavior of relaxed-stabilizer states under local bit-flip ($X$) noise.  
Let $\rho_X^{\rm rss}$ denote a relaxed-stabilizer state subjected to independent $X$-noise on each qubit with probability $p$, i.e.,
\begin{align}
&\rho_X^{\rm rss} = \bigotimes_{j=1}^N \mathcal E_{X_j}(\rho^{\rm rss}), \n
&\mathcal E_{X_j}(\rho) = (1-p)\rho + p X_j \rho X_j .
\end{align}
As derived in the Supplementary Note~5, the QFI under the $X$-noise can be expressed as
\begin{align}\label{X-noise}
\mathcal F_Q(\rho^{\rm rss}_X)&=(1-p)\sum_{\vec t:t_1=0} \rho^{\rm rss}_{\vec t,\vec t} \left(N-2\sum_{j=1}^m f(t_j,p) N_j'\right)^{2}\n 
&+p\sum_{\vec t:t_1=0} \rho^{\rm rss}_{\vec t,\vec t} \left(N-2\sum_{j=1}^m f(t_j,1\!-\!p) N_j'\right)^{2}\n 
&+4p(1-p)(N-1)
\end{align}
with $N_1' = N_1 - 1$, $N_{j \neq 1}' = N_j$, and $f(t_j,p) = p + t_j (1 - 2p)$.

In the large-$N$ limit we identify $N_j' \simeq N_j$, and define the effective minimal ratio under noise
\begin{align}
r_{\min}(p)= \min_{\vec t:t_1=0} \left(1-2\sum_{j=1}^m f(t_j,p) \xi_j  \right)^{2}.  
\end{align}
The QFI is then lower bounded by
\begin{align}
\mathcal F_Q(\rho_X^{\rm rss}) \ge \big[ (1-p) r_{\min}(p) + p \, r_{\min}(1-p) \big] N^2.
\end{align}
For the (a) biased partition with $\xi_1=(1+\delta)/2$ and $\delta>0$, one finds $r_{\min}(p) =(1-2p)^2 \delta^2$,
so that the QFI remains HL scaling up to a prefactor,
\begin{align}
\mathcal F_Q(\rho_X^{\rm rss}) \ge  (1-2p)^2 \delta^2  N^2.
\end{align}
Similarly, for the (b) uniform partition with $m$ odd, one finds $r_{\min}(p) = (1-2p)^2 / m^2$, which gives
\begin{align}
\mathcal F_Q(\rho_X^{\rm rss}) \ge (1-2p)^2 \frac{N^2}{m^2}.
\end{align}
Hence, both the biased and uniform partitions preserve HL scaling under $X$-noise, with the prefactor reduced by $(1-2p)^2$ compared to the noiseless case.  

\begin{figure}[t]
\begin{centering}
\includegraphics[scale=0.5]{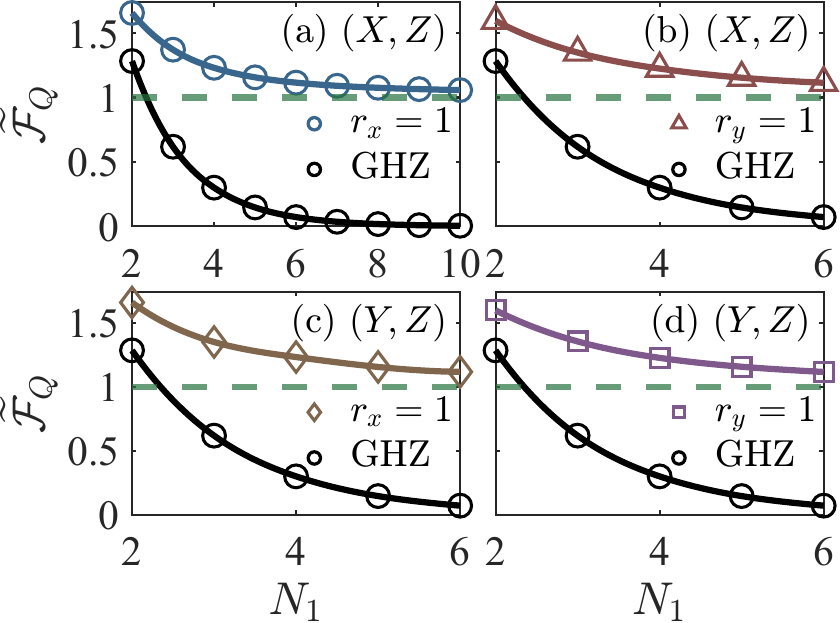}
\par\end{centering}
\caption{\label{fig5}
QFI of different probe states~\eqref{rss_qubit} with $r_x=1$, $r_y=1$, or $r_z=1$ (GHZ) under $(X,Z)$ or $(Y,Z)$ noise.
The QFI is normalized as $\widetilde{\mathcal F}_Q=\mathcal{F}_Q/[(1-2p)^2 N_1^2]$, according to Eq.~\eqref{rss_XZ}.
In all figures, $N_1=N_2$ and the noise probabilities are taken as $p=0.1$ for $X$-noise and $q=0.15$ for $Z$-noise.} 
\end{figure}

To further demonstrate the noise-resilient nature, we consider the simplest $m=2$ scenario subjected to a mixture of bit-flip ($X$) noise and phase-flip ($Z$) noise.
Specifically, qubits in the first subset~$\mathcal{N}_1$ are affected by $X$-noise with probability~$p$, while qubits in the second subset~$\mathcal{N}_2$ undergo $Z$-noise with probability~$q$. 
The resulting density matrix can be written as
\begin{align}
&\rho_{X,Z}^{\rm rss} = \left[\bigotimes_{j\in \mathcal N_1} \mathcal E_{X_j}\bigotimes_{j\in \mathcal N_2} \mathcal E_{Z_j}\right](\rho^{\rm rss}), \n
&\mathcal E_{X_j}(\rho) = (1-p)\rho + p X_j \rho X_j,\n 
&\mathcal E_{Z_j}(\rho) = (1-q)\rho + q Z_j \rho Z_j. 
\end{align}
The relaxed-stabilizer subspace~$\mathcal{R}_2$ is spanned by the two orthogonal states
\begin{align}
&\ket{\Phi}=\frac{1}{\sqrt{2}}(\ket{0}^{\otimes N_1} \ket{0}^{\otimes N_2}+\ket{1}^{\otimes N_1} \ket{1}^{\otimes N_2}),\n 
&\ket{\Psi}=\frac{1}{\sqrt{2}}(\ket{0}^{\otimes N_1} \ket{1}^{\otimes N_2}+\ket{1}^{\otimes N_1} \ket{0}^{\otimes N_2}).     
\end{align}
To simplify the expression of the relaxed-stabilizer state $\rho^{\mathrm{rss}}$, we define the following matrices on the subspace $\mathcal R_2$ 
\begin{align}
&\tau_0=\ket{\Phi}\bra{\Phi}+\ket{\Psi}\bra{\Psi},\quad \tau_y=-i\ket{\Phi}\bra{\Psi}+i\ket{\Psi}\bra{\Phi},
\n
&\tau_z=\ket{\Phi}\bra{\Phi}-\ket{\Psi}\bra{\Psi},\quad 
\tau_x=\ket{\Phi}\bra{\Psi}+\ket{\Psi}\bra{\Phi},
\end{align}
and the initial state $\rho^{\mathrm{rss}}$ can be compactly written as
\begin{align}\label{rss_qubit}
\rho^{\rm rss} = \frac{\tau_0+\vec{r}\cdot \vec \tau}{2},\quad |\vec r|\leq 1, 
\end{align}
where $\vec{r}$ is the Bloch vector and $\vec{\tau} 
= \bigl( \tau_x,\tau_y,\tau_z\bigr)$.

For the case $\vec{r} = (r_x, 0, 0)$, the QFI of the state subjected to this $(X,Z)$-noise evaluates to (see Supplementary Note~5)
\begin{align}
\mathcal F_Q(\rho_{X,Z}^{\rm rss})
&=\eta^2 N_2^2+4 N_{1} p(1-p)\frac{r_{x}^{2}+\eta^{2}-2 r_{x}^{2} \eta^{2}}{1-r_{x}^{2} \eta^{2}}\n  
& +(1-2p)^2N_{1}^{2}\frac{r_{x}^{2}+\eta^{2}-2 r_{x}^{2} \eta^{2}}{1-r_{x}^{2} \eta^{2}},
\end{align}
where $\eta=(1-2q)^{N_2}$.
For $0 < q, p < 1/2$ and in the asymptotic limit $N_1, N_2 \to +\infty$, we obtain
\begin{align}\label{rss_XZ}
\mathcal F_Q(\rho_{X,Z}^{\rm rss})=(1-2p)^2 N_1^2 r_x^2+\mathcal O(N_1). 
\end{align}
This result demonstrates that the partial HL scaling proportional to $N_1^2$ is preserved whenever $r_x\neq 0$.
For comparison, Supplementary Note~5 shows that, under the same $(X,Z)$-noise, the QFI of the GHZ state decays as $\mathcal{F}_Q = (1-2q)^{2N_2} \, g(N_1, N_2) \to 0$ for any fixed $q<1/2$ in the limit $N_2\to\infty$. 
The explicit form of $g(N_1,N_2)$ is provided therein.
Equation~\eqref{rss_XZ} therefore indicates that coherent states with $r_x\neq 0$ can outperform the GHZ state in the presence of $(X,Z)$-noise.

In addition, we have numerically examined other coherent states (e.g., with $r_y=1$) under both $(X,Z)$- and $(Y,Z)$-noise, as shown in Fig.~\ref{fig5}.
We emphasize that $X$- and $Y$-noise satisfy the Hamiltonian-not-in-Lindblad-span (HNLS) condition~\cite{Zhounc}, namely, the Hamiltonian does not commute with the corresponding noise operators in qubit systems, whereas $Z$-noise does not.
The numerical results demonstrate that coherent states remain robust under HNLS noise, as the partial Heisenberg-limit scaling proportional to $N_1^2$ is preserved under both $(X,Z)$- and $(Y,Z)$-noise (see Fig.~\ref{fig5}). Here, $\widetilde{\mathcal F}_Q$ is normalized as $\widetilde{\mathcal F}_Q=\mathcal{F}_Q/[(1-2p)^2 N_1^2]$, according to Eq.~\eqref{rss_XZ}.

\begin{figure}[t]
\begin{centering}
\includegraphics[scale=0.57]{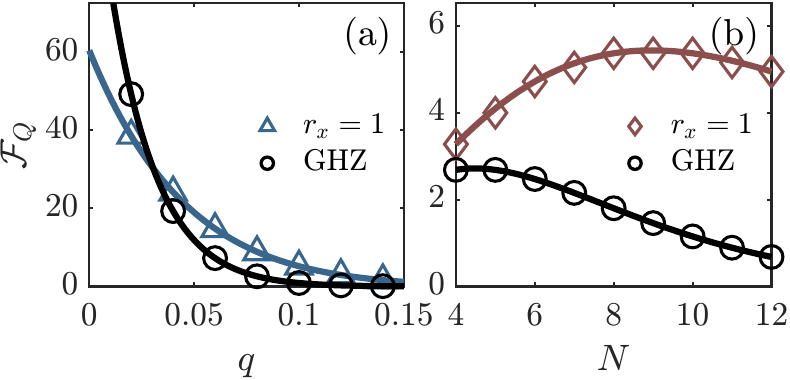}
\par\end{centering}
\caption{\label{fig6}
QFI as a function of (a) the $Z$-noise probability $q$ and (b) system size $N$ for the relaxed-stabilizer state~\eqref{rss_qubit} with $r_x=1$ and  $r_z=1$ (GHZ). 
In panel (a) $N=11$, while in panel (b) $q=0.1$.}
\end{figure}

Finally, we investigate coherent relaxed-stabilizer states subjected to $Z$-noise, where each qubit independently undergoes $Z$-noise with probability $q$.
The analytical results presented in the Supplementary Note~5 show that, for $q<1/2$ and under a uniform bipartition $N_1=N_2=N/2$, the coherent relaxed-stabilizer states~\eqref{rss_qubit} with $r_x=1$ or $r_y=1$ exhibit the QFI as $\mathcal F_Q \simeq (1-2q)^N N^2/2$,
whereas the corresponding QFI for the GHZ state behaves as $\mathcal F_Q = (1-2q)^{2N} N^2$.
The corresponding numerical results are shown in Fig.~\ref{fig6}. 
Therefore, compared to the GHZ state, the coherent relaxed-stabilizer states trade a constant factor in the noiseless precision $(q=0)$, reducing the ideal QFI from $N^2$ to $N^2/2$, for a substantially enhanced robustness against dephasing noise. 
Specifically, the exponential decay factor improves from $(1-2q)^{2N}$ to $(1-2q)^N$, thereby mitigating the noise-induced suppression of precision in the large-$N$ regime.

Although coherent relaxed-stabilizer states are already resilient to HNLS-type noise, their robustness can be further improved and generalized to arbitrary relaxed-stabilizer states, due to the underlying structure of the subspace $\mathcal R_m$.
Specifically, $\mathcal R_m$, spanned by ${\ket{\vec t}}$ in Eq.~\eqref{vect}, can be viewed as a stabilizer code associated with the stabilizer group $\mathcal K$ generated by Eqs.~\eqref{XXX} and~\eqref{ZZ}.
Since $\mathcal K$ contains $N-(m-1)$ independent stabilizer generators, this code encodes $m-1$ logical qubits.

According to stabilizer-code theory, a set of Pauli errors $\{E_\alpha\}$ is correctable on $\mathcal R_m$ if, for any $E_\alpha, E_\beta$,
\begin{align}\label{stablizer_code_error}
E_\alpha^\dagger E_\beta \notin \mathcal N(\mathcal K) \backslash \mathcal K,
\end{align}
where $\mathcal N(\mathcal K) = \{ P \in \mathcal P_N \mid P K = K P,\, \forall K \in \mathcal K \}$  is the normalizer of $\mathcal K$ in the $N$-qubit Pauli group $\mathcal P_N$.
As an example, consider errors $E_\alpha = \prod_{j=1}^m E^{(j)}$, with each block factor $E^{(j)}$ consisting of $r_j$ local $X$ errors, i.e., $E^{(j)} = \prod_{\ell=1}^{r_j} X_{j; k_\ell}$,
where $0 \le r_j \le \lfloor (N_j-1)/2 \rfloor$.
Such an error maps $\mathcal R_m$ to an orthogonal complement, satisfying the  condition~\eqref{stablizer_code_error}.
Thus each block independently corrects up to $\lfloor (N_j-1)/2 \rfloor$ bit-flip $X$-errors.

~\\

\noindent \textbf{\large{Discussion}}

\noindent 
In this work, Theorems~1 and 2 provide constructive methods for identifying optimal measurements that can saturate the QCRB based on probe-state symmetries.
Building on these results, Theorem~3 applies these constructions to the experimentally relevant scenario of local measurements, explicitly yielding the optimal local measurement strategies for parameter estimation.
By combining stabilizer theory with Theorem~3, we design estimation protocols for graph states that exploit their stabilizer structures to implement these optimal local measurements.
Furthermore, by introducing two graph-connection rules, we identify and generate additional graph states capable of achieving HL scaling under local measurements.

By reducing the number of stabilizer generators, a single graph state is enlarged into a larger state space, which we term the relaxed-stabilizer subspace.
Within certain relaxed-stabilizer subspaces, all states achieve the HL scaling in the noiseless case.
In the presence of noise, coherent states in this subspace retain HL scaling, demonstrating intrinsic noise resilience.
Finally, we identify an error-correction–based strategy to enhance their robustness against noise further.

Even though we focus on the single-parameter estimation problem, these results also provide insights for multi-parameter scenarios.
For example, in distributed quantum metrology, one aims to estimate a single parameter defined as a linear combination of multiple local parameters across different subsystems~\cite{PhysRevLett.121.043604,jkjj-3gvb,rubio2020quantum}.
In Supplementary Note~6, we show that for weakly commuting SLD operators~\cite{matsumoto2002new}, Theorem~2 can be straightforwardly generalized by replacing the single parameter $\theta$ with a vector of parameters $\vec{\theta}$, enabling the construction of optimal measurements in distributed metrology.
The same observation has been reported in continuous-variable systems~\cite{f2jf-bg7g}.
In contrast, for non-commuting SLD operators, the incompatibility of measurements generally prevents the saturation of the QCRB.
The degree of commutativity of SLDs can be quantified using the Uhlmann curvature~\cite{Carollo2018,leonforte2019uhlmann}, providing a geometric perspective on multi-parameter compatibility.
Therefore, investigating multi-parameter cases represents a promising direction for future work.

On the other hand, the graph-connection rules introduced here provide a natural framework for analysing the recently emerging problem of constructing private states in distributed quantum metrology. Such states ensure that, within a distributed quantum sensing network, only the value of a target global function can be collaboratively estimated, while any additional information about individual local parameters remains inaccessible, thereby preserving the informational privacy of the estimation process~\cite{lbfk-cykl,bugalho2025private}.
Roughly speaking, weakly connected graph structures exhibit an approximately additive behaviour, whereby the quantum Fisher information matrix becomes nearly diagonal in regimes where the diagonal elements are dominant. 
This structural property generally impedes the construction of private states for non-trivial distributed metrology tasks, namely, when the target global function depends on at least two local parameters, as it inhibits the effective suppression of locally accessible parameter information.
By contrast, strongly connected graph rules enable the construction of asymptotically private states in the large-qubit-number limit (see Supplementary Note~7 for a detailed discussion). In this regime, the twin structure plays a crucial role in suppressing local information leakage, thereby allowing the global parameter to remain accessible while preventing the extraction of individual local parameters.

Our work uncovers the deep connection between stabilizer theory and the optimal measurement problem.
We expect this approach to be applicable to a wider class of stabilizer states, such as Toric code states \cite{kitaev2003fault,PhysRevA.71.022315,PhysRevResearch.4.023144}.
Future research directions include extending our theory to continuous-variable systems \cite{PhysRevA.102.052601,PhysRevLett.128.180503}, exploring applications of Theorem~3 to qudit systems, and generalizing our results to global quantum metrology~\cite{boeyens2025role,mukhopadhyay2025current}.
We also anticipate that our theoretical schemes will find near-term applications in physical platforms such as Rydberg atoms \cite{Ding2022,RevModPhys.82.2313,qin2024scalingcomputationalorderparameters,kurzyna2025microwavefieldquantummetrologyerror}.

~\\

\noindent \textbf{\large{Methods}}
\\
\textbf{Lemma~1} Let $\rho_0 = \sum_j p_j |\psi_j\rangle\langle \psi_j|$ with $p_j\ge 0$, and let $ P = \sum_{j:p_j \neq 0} |\psi_j\rangle\langle \psi_j|$
be the projector onto the support of $\rho_0$, denoted as ${\rm Supp}(\rho_0)$.  
If $S$ is a symmetry of $\rho_0$ and anticommutes with the encoding Hamiltonian $H$ within the support of $\rho_0$, i.e., $ P \{S,H\}  P = 0$, 
then SLD operator of the parameter-encoded state $\rho_\theta = U_\theta \rho_0 U_\theta^\dagger$, can be simplified as
\begin{align}\label{M_SLD_2}
L_\theta = i [ P_\theta, H],    
\end{align}
where $ P_\theta=U_\theta  P U_\theta^\dag$, $U_\theta = e^{-i\theta H/2}$.
Moreover, 
\begin{align}\label{M_vanished_H}
\langle \psi_j | H | \psi_k \rangle = 0, \ {\rm for} |\psi_j\rangle,\ |\psi_k\rangle \in {\rm Supp}(\rho).
\end{align}

Proof: Expand the SLD operator $L_\theta$ in the eigenbasis of $\rho_\theta = \sum_j p_j |\psi_{\theta,j}\rangle\langle \psi_{\theta,j}|$, with $|\psi_{\theta,j}\rangle = U_\theta |\psi_j\rangle$
\begin{align}\label{M_SLD}
L_\theta = \sum_{p_j+p_k \neq 0} \frac{i(p_j - p_k)}{p_j + p_k} \langle \psi_j | H | \psi_k \rangle \, |\psi_{\theta,j}\rangle\langle \psi_{\theta,k}|.
\end{align}
If $|\psi_j\rangle$ and $|\psi_k\rangle$ belong to the support of $\rho_0$, then by the state symmetry $S|\psi_j\rangle = |\psi_j\rangle$ and $S|\psi_k\rangle = |\psi_k\rangle$,
\begin{align}
\langle \psi_j | H | \psi_k \rangle 
=  \langle \psi_j | P \{S,H\} P| \psi_k \rangle/2
= 0,
\end{align}
where we used \( P \{S,H\}  P = 0\).  
Hence, all terms with $p_j \neq 0$ and $p_k \neq 0$ vanish in Eq.~\eqref{M_SLD}.
The remaining nonzero contributions come from terms where exactly one of $p_j$ or $p_k$ is nonzero.  
Therefore, we have
\begin{align}
L_\theta &= -i \sum_{p_j = 0, \, p_k \neq 0} \langle \psi_j | H | \psi_k \rangle \, |\psi_{\theta,j}\rangle\langle \psi_{\theta,k}| + \mathrm{h.c.}\n 
&=-i(1- P_\theta)H P_\theta + \mathrm{h.c.}= i [ P_\theta, H].
\end{align}

\noindent
\textbf{Proof of the Theorem~1}
\\
We define the operator $\Gamma$ as $\Gamma_{jk}=[\ket{\psi_{\theta,j}}\bra{\psi_{\theta,k}},L_\theta],$
where $L_\theta$ denotes the SLD operator.
It is known that a rank-one measurement of the form $U_\theta M_x U_\theta^\dagger$ is optimal if and only if~\cite{zhou2020saturating}
\begin{eqnarray}\label{Zhou}
    \Tr{U_\theta M_xU_\theta^\dag \Gamma_{jk}}=0,
\end{eqnarray}
for all measurement outcomes $x$
and for all indices $j,k$ such that $p_j,p_k > 0$.
Since Lemma~1 is guaranteed by the state symmetry together with condition~(i), we may substitute Eq.~\eqref{M_SLD_2} into the definition of $\Gamma_{jk}$, yielding
\begin{align}\label{M_Gamma_Simplified}
 \Gamma_{jk}&=i[ \ket{\psi_{\theta,j}}\bra{\psi_{\theta,k}},[ P_\theta,H]]\n 
 &=i\{\ket{\psi_{\theta,j}}\bra{\psi_{\theta,k}},H\},
\end{align}
where Eq.~\eqref{M_vanished_H} has been used to eliminate terms of the form
$\bra{\psi_{\theta,k}} H  P_\theta$.
Substituting Eq.~\eqref{M_Gamma_Simplified} into the optimality condition~\eqref{Zhou}, we obtain
\begin{align}\label{M_THM1}
\Tr{U_\theta M_xU_\theta^\dag\Gamma_{jk}}&=i\Tr{U_\theta M_xU_\theta^\dag\{\ket{\psi_{\theta,j}}\bra{\psi_{\theta,k}},H\}}\n 
&=i\Tr{M_x\{\ket{\psi_j}\bra{\psi_k},H\}}\n 
&=i\ex{\psi_k|\{H,M_x\}|\psi_j}=0,
\end{align}
where the last equality follows directly from condition~(ii) and the fact that
$\ket{\psi_j},\ket{\psi_k}\in \mathrm{Supp}(\rho_0)$.

\noindent
\textbf{Proof of the Theorem~2}
\\
We first justify the validity of the unitary approximation. 
For a fixed true value $\theta_0$, the parametrized state $|\psi_\theta\rangle$ can be expanded to first order in $\delta_\theta = \theta - \theta_0$ as
\begin{equation}
|\psi_\theta\rangle \simeq |\psi_{\theta_0}\rangle + \delta_\theta \, |\dot\psi_{\theta_0}\rangle,
\quad 
|\dot\psi_{\theta_0}\rangle = \left. \frac{d|\psi_\theta\rangle}{d\theta} \right|_{\theta_0}.
\end{equation}
Substituting the effective Hamiltonian
\begin{equation}\label{localH}
H = 2i \big( |\dot\psi_{\theta_0}\rangle \langle \psi_{\theta_0}| - |\psi_{\theta_0}\rangle \langle \dot\psi_{\theta_0}| \big).
\end{equation}
into $|\phi_\theta\rangle \simeq e^{-i \delta_\theta H/2} |\psi_{\theta_0}\rangle$ and expanding to first order in $\delta_\theta$, we obtain
\begin{align}
|\phi_\theta\rangle &=(1-i\delta_\theta H/2)\ket{\psi_{\theta_0}}\n 
&=\ket{\psi_{\theta_0}}+\delta_\theta \ket{\dot\psi_{\theta_0}}-\delta_\theta \ex{\dot\psi_{\theta_0}|\psi_{\theta_0}}\ket{\psi_{\theta_0}}.
\end{align}
Since the expansion coefficients $\psi_j(\theta)$ are real in $\ket{\psi_\theta}=\sum_j \psi_j(\theta)\ket{v_j}$, we have $\ex{\dot\psi_\theta|\psi_\theta}=\ex{\psi_\theta|\dot\psi_\theta}$.
Furthermore, the normalization condition $d \langle \psi_\theta|\psi_\theta\rangle / d\theta = 0$ implies $\ex{\dot\psi_\theta|\psi_\theta}=0$.
Therefore, $|\phi_\theta\rangle$ coincides with $|\psi_\theta\rangle$ to first order in $\delta_\theta$, justifying the unitary approximation around $\theta_0$.

To show that $\ket{\psi_\theta}$ saturates the QCRB at $\theta_0$, it suffices to verify the conditions of Theorem~1 for the equivalent unitary-encoded state $\ket{\phi_\theta}$.  
We choose the state symmetry operator $S=\sum_j \ket{v_j}\bra{v_j}$ for $\ket{\phi_0}$ and $M_j=\ket{v_j}\bra{v_j}$ as the measurement.
Then the conditions of Theorem~1 reduce to
\begin{align}\label{conditions-thm1-method}
&P\{S,H\}P=2{\rm Re}[\ex{\psi_{\theta_0}|H|\psi_{\theta_0}}]|\psi_{\theta_0}\rangle\langle \psi_{\theta_0}|
\n 
&P\{M_j,H\}P=2\psi_j(\theta_0){\rm Re}[\ex{v_j|H|\psi_{\theta_0}}]|\psi_{\theta_0}\rangle\langle \psi_{\theta_0}|,
\end{align}
since $P=\ket{\phi_0}\bra{\phi_0}=\ket{\psi_{\theta_0}}\bra{\psi_{\theta_0}}$.
Using the definition of $H$ in Eq.~\eqref{localH} and the property $\langle \dot\psi_{\theta_0} | \psi_{\theta_0} \rangle = 0$, we have
\begin{align}
\ex{\psi_{\theta_0}|H|\psi_{\theta_0}}=2i(\ex{\dot\psi_{\theta_0}|\psi_{\theta_0}}-\ex{\psi_{\theta_0}|\dot\psi_{\theta_0}})=0,
\end{align}
and 
\begin{align}
\ex{v_j|H|\psi_{\theta_j}}&= 2i\ex{v_j|\dot\psi_{\theta_0}}=2i\dot \psi_j(\theta_0). 
\end{align}
Hence, both anti-commutators vanish, i.e.,  $P\{S,H\}P=P\{M_j,H\}P=0$, demonstrating that the conditions of Theorem~1 are satisfied.  
Consequently, the projective measurement $\{M_j\}$ saturates the QCRB for $\ket{\phi_\theta}$ and also for  $|\psi_\theta\rangle$ at $\theta_0$, completing the proof of Theorem~2.

\noindent \textbf{Lemma 2}
Let $\rho=\sum_{k=1}^s p_k \ket{\psi_k}\bra{\psi_k}$ with $p_k>0$ and consider the unitary encoding operator
$U_\theta=\exp(-i\theta H/2)$. Assume that
\begin{align}
\langle \psi_k|H|\psi_{k'}\rangle=0 \qquad \forall\, \ket{\psi_k},\ket{\psi_{k'}}\in \mathrm{Supp}(\rho).
\end{align}
Let $\{\ket{\phi_j}\}_{j=1}^s$ be an orthonormal basis of $\mathrm{Supp}(\rho)$ such that
\begin{align}
H^2\ket{\phi_j}=\lambda_j \ket{\phi_j}.
\end{align}
Then the QFI of $\rho$ with respect to $H$ satisfies
\begin{align}
\mathcal F_Q(\rho,H)=\sum_{j=1}^s \lambda_j\, \langle \phi_j|\rho|\phi_j\rangle .
\end{align}

Proof: 
For a mixed state $\rho=\sum_{k=1}^s p_k \ket{\psi_k}\bra{\psi_k}$, the QFI associated with
$U_\theta=\exp(-i\theta H/2)$ is given by \cite{liu2014quantum}
\begin{align}
\mathcal F_Q
=\sum_{k=1}^s p_k \ex{ \psi_k|H^2|\psi_k}
-\sum_{k,k'=1}^s \frac{2p_k p_{k'}}{p_k+p_{k'}}
\big|\ex{ \psi_k|H|\psi_{k'}}\big|^2 .
\end{align}
By the assumption $\ex{ \psi_k|H|\psi_{k'}}=0$ on $\mathrm{Supp}(\rho)$, the second term
vanishes identically, and hence
\begin{align}
\mathcal F_Q=\sum_{k=1}^s p_k \langle \psi_k|H^2|\psi_k\rangle .
\end{align}
Expand $\ket{\psi_k}$ in the eigenbasis $\{\ket{\phi_j}\}$ of $H^2$, $\ket{\psi_k}=\sum_{j=1}^s u_{kj}\ket{\phi_j}$, with a unitary matrix $u$. 
Using $H^2\ket{\phi_j}=\lambda_j\ket{\phi_j}$, we obtain
\begin{align}
\mathcal F_Q
&=\sum_{k=1}^s p_k \sum_{i,j=1}^s u_{ki}^* u_{kj}
\ex{ \phi_i|H^2|\phi_j} \n
&=\sum_{k=1}^s p_k \sum_{j=1}^s |u_{kj}|^2 \lambda_j
=\sum_{j=1}^s \lambda_j \sum_{k=1}^s p_k |u_{kj}|^2 .
\end{align}
Finally, noting that
\begin{align}
\ex{\phi_j|\rho|\phi_j}
=\sum_{k=1}^s p_k \ex{ \phi_j|\psi_k}\ex{ \psi_k|\phi_j}
=\sum_{k=1}^s p_k |u_{kj}|^2 ,
\end{align}
we conclude that
\begin{align}
\mathcal F_Q=\sum_{j=1}^s \lambda_j\, \ex{ \phi_j|\rho|\phi_j},
\end{align}
which completes the proof.

\noindent \textbf{ Proof of the Protocol $\textbf{2}$}

\noindent
Since the relaxed-stabilizer state $\rho^{\rm rss}$~\eqref{state_pro_2} exhibits the local-state symmetry
$X_1X_2\cdots X_N$, solving the compatibility conditions~\eqref{qubit}
implies that an optimal local estimation protocol can be constructed by choosing the Hamiltonian $H=\sum_{j=1}^N Z_j$.
It can be verified that, by definition~\eqref{vect}, the following relations hold
\begin{align}
&\ex{\vec t'|H|\vec t}=0,
\qquad t_1=t_1'=0, \n
&H^2\ket{\vec t}
=\Bigl(N-2\sum_{j=1}^m t_j N_j\Bigr)^2 \ket{\vec t}.
\end{align}
Moreover, the set
$\{\ket{\vec t}: t_1=0,\; t_{j\neq 1}\in\{0,1\}\}$
forms an orthonormal basis of $\Supp{\rho^{\rm rss}}$.
Therefore, by applying Lemma~2, the QFI is given by Eq.~\eqref{QFI_subspace}.

\bibliography{ref}

~\\
\noindent \textbf{Data availability}
\\	
\noindent
The authors declare that the data supporting the findings of this study are available within the paper and its Supplementary Information files. Should any raw data files be needed in another format they are available from the corresponding author upon reasonable request.
\\
\\
\noindent \textbf{Acknowledgments}
\\	
\noindent
We thank Jing Yang, Yun-Hao Shi and Weizhe Gao for helpful discussions.  
HLS was supported by the European Commission through the
H2020 QuantERA ERA-NET Cofund in Quantum Technologies project ``MENTA'' and received
funding under Horizon Europe programme HORIZONCL4-2022-QUANTUM-02-SGA via the
project 101113690 (PASQuanS2.1).
CW was supported by the National Research Foundation, Singapore and A*STAR under its Quantum Engineering Programme (NRF2021-QEP2-02-P03).  
SY was supported by Innovation Program for Quantum Science and Technology (2021ZD0300804).\\
\\
\\
\noindent \textbf{Author contributions} 

\noindent 
CW and SY conceived the project and designed the research.
JXL, HLS, and SY developed the theorems, constructed the protocols, and completed the proofs, and drafted the manuscript.
JXL performed the numerical calculations and prepared all figures.
HLS revised the manuscript with input from all authors.
All authors contributed to this work and reviewed the manuscript.

\noindent \textbf{Competing interests}
\\	
\noindent
The authors declare no competing interests.

\end{document}